\documentclass{article}
\usepackage{a4,amssymb}
\begin{document}
%%%%
%%%
%%% \input{tlsymbols}
\def\@psfmtname{pslatex}
\ifx\normalshape\undefined \def\normalshape{}\fi % if not pslatex define
                                                 % dummy normalshape
\ifx\fmtname\@psfmtname \else \def\cmsy@{2}\fi % make sure we can always get cmsy

\ifx\fulldiamondsuit\undefined %
\def\sometimeP{\!\,\makebox[1em]{$\sometime$}\hspace{-1em}
    \raisebox{.25ex}{\makebox[1em]{$\bullet$}}\,} %
\else %
\def\sometimeP{\!\,\raisebox{-0.2ex}{\makebox[1em]{\LARGE$\fulldiamondsuit$}}}%
\fi
        \def\sometime{\mathord{\hbox{\large$\mathchar"0\cmsy@7D$}}}

\let\was=\sometimeP

\newcommand{\always}{\raisebox{-.2ex}{
                           \mbox{\unitlength=0.9ex
                           \begin{picture}(2,2)
                           \linethickness{0.06ex}
                           \put(0,0){\line(1,0){2}}
                           \put(0,2){\line(1,0){2}}
                           \put(0,0){\line(0,1){2}}
                           \put(2,0){\line(0,1){2}}
                           \end{picture}}}
                          \,}
\newcommand{\alwaysP}{\rule[-0.2ex]{1.8ex}{1.8ex}\,}
\def\mysometime{\hspace{-1.2ex}\hbox{\large$\mathchar"0\cmsy@7D$}}
\newcommand{\next}{\!\raisebox{-.2ex}{ %possibly add a little space before
                        \mbox{\unitlength=0.9ex
                        \begin{picture}(2,2)
                        \linethickness{0.06ex}
                        \put(1,1){\circle{2}} % Draws circle with
                        \end{picture}}}       % diameter 2 at centre 1,1
                        \,}
\newcommand{\snext}{\raisebox{-.2ex}{ %possibly add a little space before
                        \mbox{\unitlength=0.9ex
                        \begin{picture}(2,2)
                        \linethickness{0.06ex}
                        \put(1,1){\circle{2}} % Draws circle with
                        % diameter 2 at centre 1,1, and puts dot in middle
                        \put(1,1){\circle*{0.4}}
                        \end{picture}}}
                        \,}
\newcommand{\slast}{\raisebox{-.2ex}{
                        \mbox{\unitlength=0.9ex
                        \begin{picture}(2,2)
                        \linethickness{0.9ex}   % which doesn't matter
                        \put(1,1){\circle{0.9}}
                        \put(1,1){\circle{1.0}}
                        \put(1,1){\circle{1.2}}
                        \put(1,1){\circle{1.4}}
                        \put(1,1){\circle{1.6}}
                        \put(1,1){\circle{1.8}}
                        \put(1,1){\circle{1.86}}
                        \put(1,1){\circle{1.92}}
                        \end{picture}}}
                        \,}
\def\until{\hbox{$\,\cal U \,$}}
\def\unless{\hbox{$\,\cal W \,$}}
\let\weaknext=\next
\let\imp=\Rightarrow
\let\iff=\Leftrightarrow
\def\ltrue{\hbox{\rm\bf true}}
\def\lfalse{\hbox{\rm\bf false}}
\newcommand{\Nat}{\mbox{\(\mathbb N\)}}
\def\lstart{\hbox{\rm\bf start}} 
\newenvironment{proof}{ \hbox{\bf Proof } }{\hfill $\square$}
\newtheorem{proposition}{Proposition}
\newtheorem{lemma}{Lemma}
\newtheorem{theorem}{Theorem}
\newtheorem{corollary}{Corollary}
%%
% \newdef{defn}{Definition}
% \newdef{example}{Example}
% \newdef{notation}{Notation}
% \newdef{remark}{Remark}
%%
\newtheorem{defn}{Definition}
\newtheorem{example}{Example}
\newtheorem{notation}{Notation}
\newtheorem{remark}{Remark}

\newcommand{\verylongrightarrow}{\raisebox{-.2ex}{
%			   \mbox{\unitlength=0.9ex
			    \begin{picture}(15,2)
			    \linethickness{0.06ex}
			    \put(-30,2){\vector(3,0){75}}
			    \end{picture}}
		       \,}
\let\prv=\vdash
\def\lland{\,\land\,}
\def\bland{\ \land\ }
\def\llor{\,\lor\,}
\newcommand{\len}{{\sf len}}
\newcommand{\const}{\hbox{\bf const}}
\newcommand{\props}{{\sf props}}
\newcommand{\ops}{{\sf ops}}
%%%

%%%%%%%%%%%%%%%%%%%%%%%%%%%%%%%%%%%%%%%%%%%%%%%%%%%%%%%%%%%%%%%%%%%

\title{Clausal Temporal Resolution} 
\author{Michael Fisher, Clare Dixon\\
        Department of Computing and Mathematics, Manchester
        Metropolitan University,\\
        Manchester M1~5GD, U.K.  
        \and
        Martin Peim\\
        Department of Computer Science, Victoria University of Manchester,\\
        Manchester M13 9PL, U.K.} 
\date{}
\maketitle
\begin{abstract}
In this article, we examine how clausal resolution can be applied to a
specific, but widely used, non-classical logic, namely discrete linear
temporal logic. Thus, we first define a normal form for temporal
formulae and show how arbitrary temporal formulae can be translated
into the normal form, while preserving satisfiability. We then
introduce novel resolution rules that can be applied to formulae in
this normal form, provide a range of examples and examine the
correctness and complexity of this approach. Finally, we describe related
work and future developments concerning this work.
\end{abstract}

%
%%%%%%%%%%%%%%%%%%%%%%%%%%%% SECTION STARTS.............
\section{Introduction}
\label{intro}
Temporal logic is a non-classical logic that was originally
developed in order to represent tense in natural
language~\cite{Pri67}. More recently, it has achieved a significant
role in the formal specification and verification of concurrent and
distributed systems~\cite{Pnu77}. It is commonly recognised that such
{\em reactive systems}~\cite{HP85} represent one of the most important
classes of systems in computer science and, although analysis of these
systems is difficult, it has been successfully tackled using modal and temporal
logics~\cite{Pnu77,emerson:90a,Stir92}.  In particular, a number of
useful concepts, such as safety, liveness and fairness can be
formally, and concisely, specified using temporal
logics~\cite{MannaPnueli92:book,emerson:90a}.

There are now a wide variety of temporal logics, differing in both
their underlying model of time (for example,
branching~\cite{EmersonSrinivasan88} versus
linear~\cite{Pnu77,MannaPnueli92:book}, and
dense~\cite{BG85} versus discrete) and their intended area
of application (for example, program
specification~\cite{MannaPnueli92:book}, temporal databases
\cite{Tan93}, knowledge representation~\cite{AF99}, executable
temporal logics~\cite{BFGOR96:book}, natural
language~\cite{Steed97}). In this paper we concentrate on a specific
but widely used temporal logic, Propositional Linear Temporal Logic (PLTL), a
discrete, linear temporal logic with finite past and infinite future;
see for example \cite{GPSS80,MannaPnueli92:book,MannaPnueli95:book}.

Given a specification of some computational system in PLTL, we may
want to establish that particular properties of the specification
hold. Thus, for concurrent systems, we must often show the absence of
deadlock, preservation of mutual exclusion, etc (see for example
\cite{Lam83spec}). There are two main approaches to temporal
verification that could be used here. If we can generate a
finite-state structure representing {\em all\/} models of the system,
then {\em model checking\/} techniques can be
applied~\cite{Holzmann97e}. Model checking involves establishing that
a specific temporal formula is satisfied in the set of models
representing the system. An alternative approach involves direct proof
in PLTL. We consider this second approach since not only may it be the
case that models are not readily available but, even if they are, many
systems we are interested in have very large, sometimes infinite,
state spaces. Importantly, the use of direct proof methods may obviate
the need to traverse all of a possible model structure.

The development of proof methods for temporal logic have followed
three main approaches: tableaux, automata and resolution. To show a
formula $\varphi$ valid, each of these methods is applied to the
negation of $\varphi$, i.e.\ $\neg \varphi$. Tableaux-based
approaches, for example \cite{Wol83,Gou84}, attempt to
systematically construct a structure from which a model can be
extracted for $\neg \varphi$. The inability to construct such a model
means that $\neg \varphi$ is unsatisfiable and therefore $\varphi$ is
valid.  The use of automata-based approaches depends on the fact that
models for PLTL are simply infinite sequences of choices for truth
values of proposition symbols. That is, an interpretation of a PLTL formula
can be viewed as an infinite word over the alphabet that is the
powerset of proposition symbols. Translations from PLTL into B{\"u}chi
Automata are given in \cite{SVW87:tcs}. If the automaton for $\neg
\varphi$ is empty then it accepts no infinite words, hence $\neg
\varphi$ is unsatisfiable and $\varphi$ is valid.

Resolution-based approaches to proof in PLTL fall into two main
classes: non-clausal and clausal. A non-clausal method described
in~\cite{AbadiManna85}, and extended to first-order temporal logic
in~\cite{AM90}, requires a large number of resolution rules,
making implementation of this method difficult. Clausal resolution was
suggested as a proof method for classical logic by
Robinson~\cite{Rob65} and was claimed to be {\em machine oriented},
i.e.\ suitable to be performed by computer as it has one rule of
inference that may be applied many times. Again, to show a formula
$\varphi$ is valid, it is negated and $\neg \varphi$ is translated
into a normal form. The resolution inference rule is applied until
either no new inferences can be made or a contradiction is
obtained. The generation of a contradiction means that $\neg \varphi$
is unsatisfiable and therefore $\varphi$ valid.

Since clausal resolution is a simple and adaptable proof method for
classical logics with a bank of research into heuristics and
strategies, it is perhaps surprising that few attempts have been made
to extend this to temporal logics. However, discrete temporal logics,
such as PLTL, are difficult to reason about as the interaction between
the $\always$-operator (meaning {\em always in the future}) and the
$\next$-operator (meaning {\em in the next moment in time}) encodes a
form of induction. Thus, a special temporal resolution rule is 
needed to handle this. There have been two previous attempts (known to
the authors) at developing clausal resolution for temporal logics. The
method described in~\cite{CaFa84} is only applicable to a subset of
the operators allowed in this paper, that is for a less expressive
language, and contains a more complex normal form. The method
described in~\cite{Ven86} is the closest to that described in this
paper, the main difference being that the reasoning is carried out
forward into the future while our approach involves reasoning
backwards until a contradiction is generated in the initial
state. Both of these are discussed further in \S\ref{related}.

The development of the new resolution method described in this paper
is motivated not only by our wish to show that such a resolution
system can be both simple and elegant, but also by our view that
clausal resolution techniques will, in the future, provide the basis
for the most efficient temporal theorem-provers. While, in previous
years, the most sucessful theorem-provers for modal and temporal
logics have been tableau-based (e.g.~\cite{Horrocks98}), the use of
resolution has now been shown to be at least
competitive~\cite{HustadtSchmidt99c}. In the classical framework,
clausal resolution has led to many refinements aimed at guiding the
search for a refutation, for example,
%\cite{WCR65,Cha70,Loveland70,Luckham70,Rob65-hyper,WCR64}. 
\cite{CL73,WOLB84}.
In
addition, several efficient, fast, and widely used resolution-based
theorem provers have been developed, for example {\sc
Otter}~\cite{Otter3.0} and {\sc Spass}~\cite{Weidenbach97}. It is our
view that a clausal temporal resolution system has the potential to
utilise a range of such efficient improvements developed for both
classical and modal resolution.

Thus, our approach is clausal. In particular, we define a very simple
(and flexible) normal form, called Separated Normal Form (SNF), that
removes all but a core set of temporal operators. Two types of
resolution rule are then defined, one analogous to the classical
resolution rule and the other a new {\em temporal resolution\/} rule.
However, due to the interaction between the $\always$ and $\next$
operators mentioned previously, the application of the temporal
resolution rule is non-trivial, requiring specialised algorithms
\cite{Dix96-CADE}. It is not our intention here to analyse
experimental results concerning use of the resolution method (which
still remain part of our future work), but simply to provide a
logically complete basis for clausal temporal resolution. While short
reports on this work have appeared previously, notably
in~\cite{Fis90-resolve}, this paper provides the first exposition of
the full completeness result for this temporal resolution method. In
addition, it provides important properties of the translation into the
normal form, and presents a simpler future-time formulation of the
method.

The structure of the paper is as follows. In \S\ref{PLTL} we give the
syntax and semantics of PLTL. In \S\ref{normal.form}, we define the
normal form (SNF), show how any PLTL formula may be translated into
SNF and consider the properties of this translation. The resolution
rules for formulae in SNF are given in \S\ref{res.rules} while example
refutations are provided in \S\ref{sec.eg}. Issues of correctness and
complexity are considered in \S\ref{sec.correct} and
\S\ref{sec.complex}, respectively. Related work is examined in
\S\ref{related} and conclusions and future work are provided in
\S\ref{sec.conc}.
%
%
%%%%%%%%%%%%%%%%%%%%%%%%%%%% SECTION STARTS.............
\section{Propositional Temporal Logic}
\label{PLTL}
Propositional Temporal Logic (PLTL) was originally developed from work
on tense logics~\cite{Pri67}, but has come to prominence through its
application in the specification and verification of both software and
hardware~\cite{Pnu77}. The particular variety of temporal logic we
consider is based on a linear, discrete model of time with finite past
and infinite future \cite{GPSS80,LPZ85}. Thus, the temporal operators
supplied operate over a sequence of distinct `moments' in time.

There are several ways to view this logic. One is as a classical
propositional logic augmented with temporal connectives (or
operators). An alternative characterisation can be given in terms of a
multi-modal language with two different modalities, one representing
the `next' moment in time, the other representing all future moments
in time (`$\next$' and `$\always$' below, respectively).

While it is possible to include past-time operators in the definition
of the logic we choose not to do so in this exposition since, as
models have a finite past, such operators add no extra expressive
power~\cite{GPSS80,LPZ85}. However, if the addition of past-time operators
makes the expression of certain properties easier (see, for example,
\cite{LPZ85}) they can be easily incorporated (see \S\ref{normal.form}
for more details).

The future-time connectives that we use include `$\sometime$' ({\em
sometime in the future}), `$\always$' ({\em always in the future}),
`$\next$' ({\em in the next moment in time}), `$\until$' ({\em
until}), and `$\unless$' ({\em unless}, or {\em weak until}). To
assist readers who may be unfamiliar with the semantics of the
temporal operators we introduce, in the next section, all operators as
basic. Alternatively we could have provided the syntax and semantics
of just a subset of the operators and introduced the remainder as
abbreviations.

\subsection{Syntax}
\label{sec.ptl.syntax}
PLTL formulae are constructed from the following elements.
% using the following connectives and proposition symbols.
%
\begin{itemize} 
\item ~A set, ${\cal P}$, of propositional symbols.
% \item ~Nullary connectives, $\ltrue$ and $\lfalse$. %and $\lstart$.  
\item ~Propositional connectives, $\ltrue$, $\lfalse$, $\neg$, $\lor$,
      $\land$, and  $\Rightarrow$.
%, and $\Leftrightarrow$.
\item ~Temporal connectives, $\next$, $\sometime$, $\always$, $\until$,
      and $\unless$. 
\end{itemize}
The set of well-formed formulae of PLTL, denoted by {\sc wff}, is
inductively defined as the smallest set satisfying the following.
\begin{itemize}
\item ~Any element of ${\cal P}$ is in {\sc wff}.
\item ~$\ltrue$ and $\lfalse$ are in {\sc wff}.
% and $\lstart$ 
% {\samepage
\item ~If $A$ and $B$ are in {\sc wff} then so are
      \begin{center}
         $\neg A$ \quad $A \lor B$ \quad $A \land B$ \quad $A
         \Rightarrow B$ 
%\quad $A \iff B$ 
\quad $\sometime A$ \quad
         $\always A$ \quad $A \until B$ \quad $A \unless B$ \quad
         $\next A.$   
      \end{center}
% }
\end{itemize}
A {\it literal} is defined as either a proposition symbol or the negation of
a proposition symbol.

\noindent An {\it eventuality} is defined as a formula of the form $\sometime A$.

%, while a {\em state formula\/} is a {\sc wff\/}
%containing no temporal operators.

\subsection{Semantics}
PLTL is interpreted over discrete, linear structures, for example
the natural numbers, $\Nat$. 
% with finite past and infinite future. 
%
A model of PLTL, $\sigma$, can be characterised as a sequence of {\em states}
$$
 \sigma = s_{0}, s_{1}, s_{2}, s_{3}, \ldots
$$
where each state, $s_{i}$, is a set of proposition symbols, representing
those proposition symbols which are satisfied in the $i^{th}$ moment in time.
As formulae in PLTL are interpreted at a particular state in the
sequence (i.e.\ at a particular moment in time), the notation
% truth of a formula $f$ is denoted by
%
$$
 (\sigma, i) \models  A
$$
denotes the truth (or otherwise) of formula $A$ in the model $\sigma$
at state index $i \in \Nat$.  For any formula $A$, model $\sigma$ and
state index $i \in \Nat$, then either $(\sigma, i) \models A$ holds or
$(\sigma, i) \models A$ does not hold, denoted by $(\sigma, i) \not
\models A$.  If there is some $\sigma$ such that $(\sigma, 0) \models
A$, then $A$ is said to be {\em satisfiable}.  If $(\sigma, 0) \models
A$ for all models, $\sigma$, then $A$ is said to be {\em valid} and is
written $\models A$. Note that formulae here are interpreted at $s_0$;
this is an alternative, but equivalent, definition to the one commonly
used~\cite{emerson:90a}.
\vspace{1em}

\noindent The semantics of {\sc wff\/} can now be given, as follows.
\begin{tabbing}
******\=**************\=******\=****************************************\kill
 \>$(\sigma, i) \models p$ \>iff \> $p ~ \in ~ s_{i}$ \qquad\quad [where 
 $p ~ \in ~ {\cal P}$]\\
 \>$(\sigma, i) \models \ltrue$\\
 \>$(\sigma, i) \not\models \lfalse$\\
 \>$(\sigma, i) \models A \land B$ \>iff \> $(\sigma, i) \models A$
 and $(\sigma, i) \models B$\\
 \>$(\sigma, i) \models A \lor B$ \>iff \> $(\sigma, i) \models A$ or
 $(\sigma, i) \models B$\\
 \>$(\sigma, i) \models A \imp B$ \>iff \> $(\sigma, i) \models \neg A$ or
 $(\sigma, i) \models B$\\
 \>$(\sigma, i) \models \neg A$ \>iff \> $(\sigma, i)\not\models A$\\ 
%
% \>$(\sigma, i) \models \lstart$ \>iff \> $i =0$\\
%
 \>$(\sigma, i) \models \next A$ \>iff \> $(\sigma, i + 1) \models A$\\
 \>$(\sigma, i) \models \sometime A$ \>iff \> there exists a $k\in\Nat$ such
 that $k \geqslant i$ and $(\sigma, k) \models A$\\
 \>$(\sigma, i) \models \always A$ \>iff \> for all $j\in\Nat$, if $j
 \geqslant i$ then $(\sigma, j) \models A$\\
 \>$(\sigma, i) \models A \until B$ \>iff \> there exists a
 $k\in\Nat$, such that $k \geqslant i$ and $(\sigma, k) \models B$\\
 \>\>\> \qquad and for all $j\in\Nat$, if $i \leqslant j < k$ then $(\sigma,
 j) \models A$\\ 
 \>$(\sigma, i) \models A \unless B$ \>iff \> $(\sigma, i)
 \models A \until B$ or $(\sigma, i) \models \always A$ 
\end{tabbing}

\subsection{Proof Theory}
The standard axioms and inference rules for PLTL are as follows (taking
the temporal operators $\next$, $\always$ and $\until$ as primitive and the
remaining as abbreviations--see \S\ref{equiv}).
%The operators $\next$, $\until$ and $\always$ are taken as primitive and
%$\sometime$ is defined as
%$$
%\sometime A \equiv \neg \always \neg A.
%$$
The axioms are all substitution instances of the following:
\begin{itemize}
\item[]
\begin{enumerate}
\item   all classical tautologies,
\item   $\prv\ \always (A \imp B ) \imp (  \always A \imp \always B)$
\item   $\prv\  \next \neg A \imp \neg \next A$
\item   $\prv\  \neg \next A \imp \next \neg A$
\item   $\prv\ \next (A \imp B) \imp (  \next A \imp \next B)$
\item   $\prv\ \always A \imp A \wedge \next \always A$    
\item   $\prv\ \always( A \imp \next A) \imp ( A \imp \always A)$
\item  $\prv\ (A \until B ) \imp \sometime B$
\item   $\prv\ (A \until B ) \imp (B \vee ( A 
               \wedge \next (A \until B  )))$
\item   $\prv\  (B \vee ( A 
               \wedge \next (A \until B  )))\imp (A \until B )$
\end{enumerate}
\end{itemize}
The inference rules are modus ponens 
$$
\frac{\prv A\ \ \prv\,A \imp B }{\prv B}
$$
and 
generalization
$$
\frac{\prv A}{\prv\,\always A}\,\, .
$$
\begin{theorem}\cite{GPSS80} (Soundness)
If $\vdash A$ then $A$ is valid in PLTL.
\end{theorem}
\begin{theorem}\cite{GPSS80}(Completeness)
If $A$ is valid in PLTL then $\vdash A$.
\end{theorem}

A complete axiom system for PLTL with future-time temporal operators is
given in \cite{GPSS80}. The axiom system presented here is slightly
different from the original due to slight differences in the semantics of the
connectives used. We note that it is difficult to use such an axiom system for
automated theorem proving as it is not always clear which step should be taken next
to move towards a proof.

\subsubsection{Some Equivalences}
\label{equiv}
To assist the understanding of the translation to the normal form given
in \S\ref{normal.form} we list some equivalent PLTL formulae.
$$
\begin{array}{rcl}
\next(A \land B) & \equiv & \next A \land \next B \\
%\end{array}
%$$
%$$
%\begin{array}{rcl}
\neg \next A & \equiv & \next \neg A \\
\always A & \equiv & A \land \next \always A \\
\sometime A & \equiv & A \lor \next \sometime A \\
\neg \always A & \equiv & \sometime \neg A \\
(A \until B) & \equiv & B \lor (A \land \next(A \until B)) \\
(A \until B) & \equiv & (A \unless B) \land \sometime B \\
\neg (A \until B) & \equiv & \neg B \unless (\neg A \land \neg B) \\
(A \unless B) & \equiv & B \lor (A \land \next(A \unless B)) \\
\neg (A \unless B) & \equiv & \neg B \until (\neg A \land \neg B) \\
\end{array}
$$
These are standard and are given in \cite{Gou84} for example.
%%%%%%%%%%%%%%%%%%%%%%%%%%%% SECTION STARTS.............
\section{A Normal Form for Propositional Temporal Logic}
\label{normal.form}
\subsection{Separated Normal Form}
\label{sub.snf}
The resolution method is clausal, and so works on formulae transformed
into a normal form. The normal form, called Separated Normal Form
(SNF), was inspired by (but does not require) Gabbay's separation
result \cite{Gab89}, which states that temporal formulae can be
transformed into their past, present and future-time components. The
normal form we present comprises formulae that are implications with
present-time formulae on the left-hand side and (present or)
future-time formulae on the right-hand side.  The transformation into
the normal form reduces most of the temporal operators to a core set
and rewrites formulae to be in a particular form. The transformation
into SNF depends on three main operations: the renaming of complex
subformulae; the removal of temporal operators; and classical style
rewrite operations.

Renaming, as suggested in~\cite{PlaistedGreenbaum86}, is a way of
preserving the structure of a formula when translating into a normal
form in classical logic. Here, complex subformulae can be replaced by
a new proposition symbol and the truth value of the new proposition symbol is linked
to the subformula it represents at all points in time. The removal of
temporal operators is carried out by using (fixed point) equivalences,
for example
$$
\always p \equiv (p \land \next \always p)
$$
that `unwind' the temporal operators to give formulae that need to
hold both now and in the future. Classical rewrite operations allow us
to manipulate formulae into the required form.

To assist in the definition of the normal form we introduce a further
(nullary) connective $\lstart$, that holds only at the beginning of time,
i.e.\ 
$$
 (\sigma, i) \models \lstart~~~~ \mbox{ iff }~~~~ i =0.
$$
This allows the general form of the (PLTL-clauses of the) normal form to be implications. An
alternative would be to allow disjunctions of literals as part of the
normal form representing the clauses holding at the beginning of time.
\vspace{1em}

\noindent Formulae in SNF are of the general form
 $$
 \always\bigwedge_i A_i
 $$
where each $A_i$ is known as a {\em PLTL-clause\/} (analogous to a `clause'
in classical logic)
and must be one of the following forms with each particular $k_a$,
$k_b$, $l_c$, $l_d$ and $l$ representing a literal.
$$
\begin{array}{lcll}
\lstart    &\imp & \displaystyle\bigvee_{c} l_c
	 \qquad& \hbox{\rm (an {\em initial} PLTL-clause)}\\[4ex]
\displaystyle\bigwedge_{a} k_a        &\imp 
	 & \next \displaystyle\bigvee_{d} l_d
		 & \hbox{\rm (a {\em step} PLTL-clause)}\\[4ex]
\displaystyle\bigwedge_{b} k_b        &\imp & \sometime l
		 & \hbox{\rm (a {\em sometime} PLTL-clause)}
% \qquad\fbox{Shall
%	 we use $\sometime\displaystyle\bigvee_{e} l_e$?}
\end{array}
$$
%
% Here $k_a$, $l_b$, and $l$ are literals. 
For convenience, the outer `$\!\always$' and `$\land$' connectives are
usually omitted, and the set of PLTL-clauses $\{A_i\}$ is considered.
Different variants of the normal form have been
suggested~\cite{Fisher91normalizing,FisherNoel90syn,Fisher97:JLC}. For
example, where PLTL is extended to allow past-time operators the normal
form has $\lstart$ or $\slast A$ (where `$\slast$' means in the {\em
previous moment} in time and $A$ is a conjunction of literals) on the
left-hand side of the PLTL-clauses and a present-time formula or eventuality
(i.e.\ `$\sometime l$') on the right-hand side. Other versions allow
PLTL-clauses of the form $\lstart \imp \sometime l$.  These are all
expressively equivalent when models with finite past are considered.

To apply the temporal resolution rule (see \S\ref{tresrule:old}), one
or more step PLTL-clauses may need to be combined. Consequently, a variant on
SNF called {\em merged-SNF (SNF$_m$)} \cite{Fis90-resolve}, is also defined.
Given a set of PLTL-clauses in SNF, any PLTL-clause in SNF is also a
PLTL-clause in SNF$_{m}$. Any two PLTL-clauses in SNF$_{m}$
may be combined to produce a PLTL-clause in SNF$_{m}$ as follows. 
% where $C\land D$ is rewritten into DNF (to maintain the basic PLTL-clause
% structure). 
%
$$
\begin{array}{rcl}
	      A & \Rightarrow & \next C \\
	      B & \Rightarrow & \next D \\ \hline
	       (A \land B) & \Rightarrow & \next (C \land D )
\end{array}
$$
Thus, any possible conjunctive combination of SNF PLTL-clauses
can be represented in SNF$_m$.
\subsection{Translation into SNF}
\label{tran.to.snf}
In this section, we review the translation of an arbitrary PLTL formula
into the normal form (this extends the exposition provided
in~\cite{Fisher97:JLC}). The procedure uses the technique of renaming 
complex subformulae by a new proposition symbol and the truth value of the new
proposition symbol is linked to that of the renamed formula at all moments in time.
Thus, in the exposition below the new proposition symbols introduced, namely those
indicated by ${\bf v}$, ${\bf y}$ and ${\bf z}$ must be new at each iteration of
the procedure. In the remainder of \S\ref{normal.form} we show such  new
proposition symbols in bold face type. 

Take any formula $A$ of PLTL and translate into SNF by applying the
$\tau_0$ and $\tau_1$ transformations described below
(where ${\bf y}$ is a new proposition symbol).
$$
\begin{array}{rcl}
\tau_0[A] & \longrightarrow  &
\always(\lstart  \imp  {\bf y}) \land \tau_1[\always({\bf y}  \imp  A)]
\end{array}
$$
Next, we give the $\tau_1$ transformation where $x$ is a proposition symbol.
If the main operator on the right of the implication is a classical
operator (other than non-negated disjunction) remove it as follows. 
$$
\begin{array}{rcl}
\tau_1[\always(x \imp (A \land B))] & \longrightarrow  &
\tau_1[\always(x  \imp  A)] \land 
\tau_1[\always(x  \imp  B)]\\
\\
\tau_1[\always(x \imp (A \imp B))]
 & \longrightarrow  &
\tau_1[\always(x  \imp  (\neg A \lor B))]\\
\\
% \tau_1[\always(x \imp (A \iff B))]
% & \longrightarrow  &
%\tau_1[\always(x  \imp  (\neg A \lor B))] \land
%\tau_1[\always(x  \imp  (\neg B \lor A))]\\
%\\
\tau_1[\always(x \imp \neg (A \land B))]
 & \longrightarrow  &
\tau_1[\always(x  \imp  (\neg A \lor \neg B))]\\
\\
\tau_1[\always(x \imp \neg (A \imp B))]
 & \longrightarrow  &
\tau_1[\always(x  \imp  A)] \land 
\tau_1[\always(x  \imp  \neg B)]\\
\\
%\end{array}
%$$
%$$
%\begin{array}{rcl}
\tau_1[\always(x \imp \neg (A \lor B))]
 & \longrightarrow  &
\tau_1[\always(x  \imp  \neg A)] \land
\tau_1[\always(x  \imp  \neg B)]\\
\\
%\tau_1[\always(x \imp \neg (A \iff B))]
% & \longrightarrow  &
%\tau_1[\always(x  \imp  (A \lor B))] \land \tau_1[\always(x \imp (\neg A \lor \neg B))]\\
\end{array}
$$
Complex subformulae enclosed in any temporal operators are renamed as
follows (where ${\bf v}$, ${\bf y}$ and ${\bf z}$ are new proposition symbols).
$$
\begin{array}{rcll}
&&&\mbox{$A$ neither literal}\\
\tau_1[\always(x \imp \next A )]
 & \longrightarrow  &
\always(x  \imp  \next {\bf y}) \land
\tau_1[\always({\bf y}  \imp  A)]
&
\mbox{nor disjunction}\\
&&&\mbox{of literals.}\\
% \\
\tau_1[\always(x \imp \neg \next A)]
 & \longrightarrow  &
\always(x  \imp  \next {\bf y}) \land
\tau_1[\always({\bf y}  \imp  \neg A)] \\
\\
\tau_1[\always(x \imp \always A)]
 & \longrightarrow  &
\tau_1[\always(x  \imp  \always {\bf y})] \land
\tau_1[\always({\bf y}  \imp  A)]
 &\mbox{$A$ not a literal.}\\
\\
\tau_1[\always(x \imp \neg \always A)]
 & \longrightarrow  &
\always(x  \imp  \sometime {\bf y}) \land
\tau_1[\always({\bf y}  \imp  \neg A)]\\
\\
\tau_1[\always(x \imp \sometime A )]
 & \longrightarrow  &
\always(x  \imp  \sometime {\bf y}) \land 
\tau_1[\always({\bf y}  \imp  A)]
 &\mbox{$A$ not a literal.}\\
\\
\tau_1[\always(x \imp \neg \sometime A)]
 & \longrightarrow  &
\tau_1[\always(x  \imp  \always {\bf y})] \land
\tau_1[\always({\bf y}  \imp  \neg A)]\\
\end{array}
$$
$$
\begin{array}{rcll}
% \end{array}
% $$
% $$
% \begin{array}{rcll}
% \\
\tau_1[\always(x \imp  A \until B)]
 & \longrightarrow  &
\tau_1[\always(x  \imp  {\bf y} \until B)] \land
\tau_1[\always({\bf y}  \imp  A)]
 &\mbox{$A$ not a literal.}\\
\\
\tau_1[\always(x \imp  A \until B)]
 & \longrightarrow  &
\tau_1[\always(x  \imp  A \until {\bf y})] \land 
\tau_1[\always({\bf y}  \imp  B)]
 &\mbox{$B$ not a literal.}\\
\\
\tau_1[\always(x \imp \neg (A \until B))]
 & \longrightarrow  &
\tau_1[\always(x  \imp  ({\bf y} \unless {\bf v}))] \land \tau_1[\always({\bf y} \imp \neg B)]\,\land \\
& &  \tau_1[\always({\bf v} \imp ({\bf y} \land {\bf z}))] \land \tau_1[\always({\bf z} \imp \neg A)]\\
\\
\tau_1[\always(x \imp  A \unless B)]
 & \longrightarrow  &
\tau_1[\always(x  \imp  {\bf y} \unless B)] \land
\tau_1[\always({\bf y}  \imp  A)]
 &\mbox{$A$ not a literal.}\\
% \end{array}
% $$
% $$
% \begin{array}{rcll}
\\
\tau_1[\always(x \imp  A \unless B)]
 & \longrightarrow  &
\tau_1[\always(x  \imp  A \unless {\bf y})] \land
\tau_1[\always({\bf y}  \imp  B)]
 &\mbox{$B$ not a literal.}\\
\\
\tau_1[\always(x \imp\!\neg (A \unless B))]
 & \longrightarrow  &
\tau_1[\always(x  \imp  ({\bf y} \until {\bf v}))] \land \tau_1[\always({\bf y} \imp \neg B)] \,\land \\
& & \tau_1[\always({\bf v} \imp ({\bf y} \land {\bf z}))] \land \tau_1[\always({\bf z} \imp \neg A)]\\
\end{array}
$$
The negated $\unless$ and $\until$ operators involve the introduction of
three new proposition symbols. Consider the transformation applied to $x \imp
\neg(A \until B)$. Applying the equivalence provided in \S\ref{equiv} we
have $x \imp (\neg B \unless (\neg A \land \neg B))$. To avoid repeating
the subformula $\neg B$ in the translation, and so that the resultant
$unless$ operator is applied to proposition symbols we introduce three new
variables, ${\bf y}$ replaces $\neg B$, ${\bf z}$ replaces $\neg A$, ${\bf v}$ replaces
${\bf y} \land {\bf z}$.

Then, any temporal operators, applied to literals, that are not allowed
in the normal form are removed as follows (where, again, ${\bf y}$ is a new
proposition symbol and $l$ and $m$ are literals).
$$
\begin{array}{rcll}
\tau_1[\always(x \imp  \always l )]
 & \longrightarrow  &
\begin{array}{rcl} 
\tau_1[\always(x & \imp & l)] \; \land  \\
\tau_1[\always(x & \imp & {\bf y})] \; \land \\
\always({\bf y} & \imp & \next l) \; \land \\
\always({\bf y} & \imp & \next {\bf y}) 
\end{array} \\
% &\mbox{$l$ is a literal.}\\
\\
\tau_1[\always(x \imp  l \until m)]
 & \longrightarrow  &
\begin{array}{rcl} 
\always(x & \imp  & \sometime m ) \; \land \\
\tau_1[\always(x & \imp & (l \lor m))] \; \land \\
\tau_1[\always(x & \imp & ({\bf y} \lor m))]\; \land \\
\always({\bf y} & \imp & \next(l \lor m)) \; \land \\
\always({\bf y} & \imp & \next ({\bf y} \lor m)) 
\end{array}  \\
% &\mbox{$l$, $m$ are literals.}\\
\\
\tau_1[\always(x \imp  l \unless m)]
 & \longrightarrow  &
\begin{array}{rcl} 
\tau_1[\always(x & \imp & (l \lor m))]\; \land \\
\tau_1[\always(x & \imp & ({\bf y} \lor m))]\; \land \\
\always({\bf y} & \imp & \next(l \lor m)) \; \land \\
\always({\bf y} & \imp & \next ({\bf y} \lor m)) 
\end{array} \\
% &\mbox{$l$, $m$ are literals.}\\
\end{array}
$$
Next, we use renaming on formulae whose right-hand side has
disjunction as its main operator but may not be in the correct form,
where ${\bf y}$ is a new proposition symbol, $D$ is a disjunction of
formulae and $A$ is neither a literal nor a disjunction of literals.

$$
\begin{array}{rcll}
\tau_1[\always(x \imp D \lor A )]
 & \longrightarrow  &
\begin{array}{rcl} 
\tau_1[\always(x & \imp & D \lor {\bf y})] \ \,\land \\
\tau_1[\always({\bf y} & \imp & A)]
\end{array} \\
%& 
%\quad\begin{array}{c}
%\mbox{$A$ is neither a literal nor}\\
%\mbox{a disjunction of literals.}\\
%\end{array}\\
\end{array}
$$
Finally, we rewrite formulae, containing no temporal operators, whose
right-hand side is a disjunction of literals, $\ltrue$ or $\lfalse$
(note that $\neg \ltrue$ and $\neg \lfalse$ are rewritten to $\lfalse$
and $\ltrue$ respectively) into PLTL-clause form and stop applying the
transformation to PLTL-clauses already in the correct form (where
$D$ is a literal or disjunction of literals and $l$ and each $l_i$ are
literals). 
$$
\begin{array}{rcll}
\tau_1[\always(x \imp D  )]
 & \longrightarrow  &
\begin{array}{rcl} 
\,\,\always(\lstart  & \imp & \neg x \lor D) \; \land \\
\always(\ltrue  & \imp & \next(\neg x \lor D)) 
\end{array} \\
%& 
%\begin{array}{c}
%\mbox{$D$ is a literal or}\\
%\mbox{disjunction of literals.}\\
%\end{array}\\
\\
\tau_1[ \always(x \imp \ltrue  )]
 & \longrightarrow  &
\begin{array}{rcl} 
\,\,\always(\lstart  & \imp & \ltrue) \; \land \\
\always(\ltrue  & \imp & \next\ltrue) 
\end{array} & \\
\\
%% \tau_1[\always(x \imp \neg \ltrue )]
%% & \longrightarrow  &
%% \begin{array}{rcl} 
%% \,\,\always(\lstart  & \imp & \neg x) \; \land \\
%% \always(\ltrue  & \imp & \next\neg x) 
%% \end{array} & \\
%% \\
\tau_1[\always(x \imp \lfalse  )]
 & \longrightarrow  &
\begin{array}{rcl} 
\,\,\always(\lstart  & \imp & \neg x) \; \land \\
\always(\ltrue  & \imp & \next\neg x) 
\end{array} & \\
\\
%% \tau_1[\always(x \imp \neg \lfalse  )]
%%  & \longrightarrow  &
%% \begin{array}{rcl} 
%% \,\,\always(\lstart  & \imp & \ltrue) \; \land \\
%% \always(\ltrue  & \imp & \next\ltrue) 
%% \end{array} & \\
%% \\
\tau_1[\always(x \imp \sometime l  )]
 & \longrightarrow  &
\,\,\always(x  \imp \sometime l) \\
% & \mbox{$l$ is a literal.}\\
\\
\tau_1[\always(x \imp \next(l_1 \lor \ldots \lor l_n ))]
 & \longrightarrow  &
\,\,\always(x  \imp \next(l_1 \lor \ldots \lor l_n )) \\
% & \mbox{each $l_i$ is a literal.}\\
\end{array}
$$
Thus, the above transformations are applied until the formula is in
the form 
$$
\bigwedge_i \always A_i
$$
where each $A_i$ is one of the three required formats. This, in turn,
is equivalent to
$$
\always \bigwedge_i  A_i.
$$
% See~\cite{Fisher97:JLC} for further details. 

\subsection{Properties of the Translation to SNF}
\label{snf.props}
% \subsubsection{Preservation of satisfiability}
Our aim is to show that the transformation is satisfiability preserving. 
This is shown in two parts. Firstly any model for a transformed formula is
also a model for the original and secondly given a model for a PLTL formula
there is always a model for its transformation into the normal form.

Thus firstly, we show that
$$
\models \tau_0[W] \imp W
$$
i.e.\ any model for the transformed formula is a model for the original.
However before we show this we first prove a lemma.
\begin{lemma}
\label{sigma1}
For all PLTL formulae $W$
$$
\models \tau_1[\always(x \imp W)] \imp \always (x \imp W)
$$
where $x$ is a proposition symbol.
\end{lemma}
\begin{proof}
The proof is carried out by induction on the structure of $W$.
For the base cases we have the following.
$$
\begin{array}{lrcl}
1. \qquad\qquad & \tau_1[\always(x \imp \sometime l)] & = & \always (x \imp \sometime l)\\
2. & \tau_1[\always(x \imp l_1 \lor \ldots \lor \ l_n)] & = & \always (\lstart \imp
\neg x \lor l_1 \lor \ldots \lor l_n) \land \\
& & & \always(\ltrue \imp \next (\neg x \lor l_1 \lor \ldots \lor l_n)) \\ 
& & \imp & \always (x \imp (l_1 \lor \ldots \lor l_n))\\
3. & \tau_1[\always(x \imp \ltrue  )] & = & \always(\lstart   \imp  \ltrue) \;
\land  \\
 &&&               \always(\ltrue   \imp  \next\ltrue)  \\
 &&\imp & \always(x \imp \ltrue) \\
%
%% 4. & \tau_1[\always(x \imp \neg \ltrue  )] & \imp & \always(\lstart   \imp  \neg x) \;
%% \land  \\
%% &&&                \always(\ltrue   \imp  \next\neg x)  \\
%% &&& \always(x \imp \neg \ltrue) \\
4. & \tau_1[\always(x \imp \lfalse  )] & = & \always(\lstart   \imp  \neg x) \;
\land  \\
&&&                \always(\ltrue   \imp  \next\neg x)  \\
&&\imp& \always(x \imp \lfalse) \\
%
%% 6. & \tau_1[\always(x \imp \neg \lfalse )] & \imp & \always(\lstart \imp \ltrue) \;
%% \land  \\
%% &&&                \always(\ltrue   \imp  \next\ltrue)  \\
%% &&& \always(x \imp \neg \lfalse) \\
%
5. & \tau_1[\always(x \imp \next(l_1 \lor \ldots \lor \ l_n))] & = & \always (x \imp
\next(l_1 \lor \ldots \lor l_n))\\
\end{array}
$$
Now, we assume that the lemma holds for $A$, $B$, $\neg A$ and $\neg
B$, e.g. $\tau_1[\always(x \imp A)] \imp \always (x \imp A)$, and show
it holds for all combinations of operators or negated operators, e.g.\
$A \land B$, $\neg (A \land B)$, $\always A$, $\lnot\always A$.  We
consider the cases for $\always A$, $\neg \always A$, $A \unless B$ and
$\neg (A \unless B)$ and note that proofs for the other operators are
similar (where ${\bf v}$, ${\bf w}$, ${\bf y}$ and ${\bf z}$ are new
proposition symbols). 
%  refer the reader to Appendix~\ref{proofs} for the remaining cases.
%
$$
\begin{array}{rcl} 
\tau_1[\always(x \imp \always A)] & = & \tau_1[\always( x \imp \always {\bf y})]
                                 \land  \tau_1[\always( {\bf y} \imp A)]\\
                                  & = & \tau_1[\always( x \imp {\bf y})] \land
                                        \tau_1[\always( x \imp {\bf z})] \land
                                               \always( {\bf z} \imp \next {\bf y}) \land\\
                                  &   & \ \ \always( {\bf z} \imp \next {\bf z})\land
                                        \tau_1[\always( {\bf y} \imp A)] \\
                                  & \imp & \always(\lstart \imp \neg x \lor {\bf y})
                                 \land \always(\ltrue \imp \next(\neg x \lor {\bf y})) \land \\
                        &&         \always(\lstart \imp \neg x \lor {\bf z})
                                   \land \always(\ltrue \imp \next(\neg x \lor {\bf z})) \land\\
                               && \always({\bf z} \imp \next {\bf y})
                                 \land \always({\bf z} \imp \next {\bf z})
                               \land \always({\bf y} \imp A)\\
                              & \imp & \always( x \imp \always A)\\
\end{array}
$$
where  $\tau_1[\always({\bf y} \imp  A)] \imp \always({\bf y} \imp A)$ from the
induction hypothesis.
$$
\begin{array}{rcl}

\tau_1[\always(x \imp \neg \always A)] & = & \always( x \imp \sometime {\bf y}) 
                               \land  \tau_1[\always({\bf y} \imp \neg A)]\\
                              & \imp & \always( x \imp \sometime {\bf y}) \land
                               \always({\bf y} \imp \neg A)\\
                              & \imp & \always( x \imp \sometime \neg A)\\
                              & \imp & \always( x \imp \neg \always A)\\
\end{array}
$$
where $\tau_1[\always({\bf y} \imp \lnot A)] \imp \always({\bf y} \imp \lnot A)$
 from the induction hypothesis.
$$
\begin{array}{rcl}
\tau_1[\always(x \imp (A \unless B))] & = & \tau_1[\always( x \imp {\bf y} \unless {\bf z})] \land
                                  \tau_1[\always({\bf y} \imp A)] \land 
                                   \tau_1[\always({\bf z} \imp B)] \\
                             & = &  \tau_1[\always( x \imp {\bf y} \lor {\bf z})] \land
                                   \tau_1[\always( x \imp {\bf w} \lor {\bf z})] \land \\
                             &&     \always( {\bf w} \imp \next({\bf y} \lor {\bf z})) \land
                                   \always( {\bf w} \imp \next({\bf w} \lor {\bf z})) \land\\
                              &&   \tau_1[\always({\bf y} \imp A)] \land
                                   \tau_1[\always({\bf z} \imp B)]  \\
                    & \imp & \always( \lstart \imp \neg x \lor {\bf y} \lor {\bf z}) \land
                        \always( \ltrue \imp \next(\neg x \lor {\bf y} \lor {\bf z})) \land\\
                       &&    \always( \lstart \imp \neg x \lor {\bf w} \lor {\bf z})\land
                           \always( \ltrue \imp \next(\neg x \lor {\bf w} \lor {\bf z})) \land \\
                    &&              \always( {\bf w} \imp \next({\bf y} \lor {\bf z})) \land
                                   \always( {\bf w}\imp \next({\bf w} \lor {\bf z})) \land \\
                   &&               \always({\bf y} \imp A) \land
                                   \always({\bf z} \imp  B) \\
                   & \imp & \always(x \imp (A \unless B))\\
\end{array}
$$
$$
\begin{array}{rcl}
 
\tau_1[\always(x \imp \neg (A \unless B))] 
                             & = & \tau_1[\always( x \imp ({\bf y} \until {\bf v})] \land
                                   \tau_1[\always({\bf v} \imp ({\bf y} \land {\bf z}))] \land
                                   \tau_1[\always({\bf y} \imp \neg B)] \land \\
                             &&    \tau_1[\always({\bf z} \imp \neg A)] \\
                             & = & \tau_1[\always( x \imp {\bf v} \lor {\bf y})] \land
                                   \tau_1[\always( x \imp {\bf v} \lor {\bf w})] \land 
                                   \always(x \imp \sometime  {\bf v}) \land \\
                             &&     \always( {\bf w} \imp \next({\bf v} \lor {\bf y})) \land
                                   \always( {\bf w} \imp \next({\bf v} \lor {\bf w})) \land\\
                              &&      \tau_1[\always({\bf v} \imp ({\bf y} \land {\bf z}))] \land
                                    \tau_1[\always({\bf y} \imp \neg B)] \land
                                   \tau_1[\always({\bf z} \imp (\neg A))]  \\
                        & \imp & \always( \lstart \imp \neg x \lor {\bf v} \lor {\bf y}) \land
                        \always( \ltrue \imp \next(\neg x \lor {\bf v} \lor {\bf y})) \land\\
                       &&    \always( \lstart \imp \neg x \lor {\bf v} \lor {\bf w})\land
                        \always( \ltrue \imp \next(\neg x \lor {\bf v} \lor {\bf w})) \land\\
     &&                    \always(x \imp \sometime  {\bf v}) \land \\
                    &&              \always( {\bf w} \imp \next ({\bf v} \lor {\bf y})) \land
                                   \always( {\bf w}\imp \next({\bf v} \lor {\bf w})) \land \\
                     &&    \always(\lstart \imp \neg {\bf v} \lor {\bf y}) \land
                         \always(\lstart \imp \neg {\bf v} \lor {\bf z}) \land \\
                     &&    \always(\ltrue \imp \next (\neg {\bf v} \lor {\bf y})) \land
                         \always(\ltrue \imp \next(\neg {\bf v} \lor {\bf z})) \land \\
        &&             \always({\bf y} \imp \neg B) \land
                                   \always({\bf z} \imp (\neg A)) \\
                   & \imp & \always(x \imp ((\neg B) \unless (\neg A
\land \neg B))) \land \always (x \imp \sometime (\neg A \land \neg B))\\
         & \imp & \always(x \imp ((\neg B) \until (\neg A \land \neg B)))\\

         & \imp & \always(x \imp \neg (A \unless B))\\
\end{array}
$$
%$$
%\begin{array}{rcl}
% \\
\end{proof}

\begin{lemma}
\label{sigma0}
For all PLTL formulae $W$
$$
\models \tau_0[W] \imp W
$$
\end{lemma}
\begin{proof}
For any PLTL formula $W$, the first step in the transformation is to
anchor $W$ to the first moment in time, i.e.\ $\tau_0[W]
\longrightarrow \always(\lstart \imp x) \land \tau_1[\always(x \imp
W)]$. From Lemma~\ref{sigma1} we have shown that $\tau_1[\always(x
\imp W)] \imp \always(x \imp W)$. Thus, as $x$ holds at the first
moment in time and the transformation implies that $(x \imp W)$ holds
at every moment in time, then $W$ also holds now.
\end{proof}
\vspace{2ex}

Next we show that for any satisfiable formula its translation is also
satisfiable, i.e.\ for any PLTL formula $W$, if $W$ is satisfiable then
$\tau_0[W]$ is satisfiable. This is established by showing that given a
model for a formula at some stage in the 
transformation process
for each step carried out in the
transformation we can find a model for the transformed formula.
\begin{defn}{[Pre-PLTL-clause form]} 
A PLTL formula is said to be in {\em pre-PLTL-clause from\/} if, and only if,
it has the structure
$$
(x_i \imp W_i)
$$
where $x_i$ is a proposition symbol (or $\lstart$) and $W_i$ is a PLTL formula.
\end{defn}
% First we prove the following lemma.
\begin{lemma}
\label{lem.one.step}
Let $\sigma$ be a model such that 
$$(\sigma,0) \models \left[\bigwedge_h \always R_h\right] \land
 \always(x \imp W)$$
where each $R_h$ is in pre-PLTL-clause form (i.e.\ an implication where the
proposition symbol on the left hand side of each implication may be different). Then, there
exists a model $\sigma'$ such that
$$(\sigma',0) \models\left[ \bigwedge_h \always R_h\right] \land
\bigwedge_j \always S_j \land \bigwedge_k \always T_k$$
where $R_h$ is in pre-PLTL-clause form, $S_j$ is in pre-PLTL-clause form and $T_k$
is in PLTL-clause form resulting from one step of the $\tau_1$
transformation, i.e.
$$
\tau_1[\always (x \imp W)] \longrightarrow \left[\bigwedge_j
\tau_1[\always S_j]\right] \land \bigwedge_k \always T_k.
$$
\end{lemma}
\begin{proof}
We examine the structure of $W$. There are three main types of
transformation that can be applied: the removal of classical
operators, the renaming of complex subformulae and the rewriting of
temporal operators applied to literals. We begin by considering the
removal of classical operators.

First, assume $W$ is a conjunction $A \land B$, i.e.
$$(\sigma,0) \models \left[\bigwedge_h \always R_h\right] \land
\always(x \imp (A \land B)).$$
Applying the $\tau_1$ translation we have
$$\tau_1[\always(x \imp (A \land B))]
 \longrightarrow  \tau_1[\always(x  \imp  A)] \land  \tau_1[\always(x
 \imp  B)] 
$$
and so we must show there is a model $\sigma'$ such that 
$$(\sigma',0) \models \left[\bigwedge_h \always R_h\right] \land
\always(x \imp A) \land \always(x \imp B).$$
Now, as $(\sigma,0) \models \always(x \imp (A \land B))$ for all $i
\in \Nat$, then if $(\sigma,i)\models x$ both $(\sigma,i) \models A$
and $(\sigma,i) \models B$. That is 
$$(\sigma,0) \models \left[\bigwedge_h \always R_h\right] \land
\always(x \imp A) \land \always(x \imp B).$$
So, by setting $\sigma'$ equal to $\sigma$ we have such a model. The
proofs are similar for the other classical logic operators.

Next, we consider renaming transformations and assume $W$ is of the
form $\always A$ where $A$ is not a literal. Now, assume that there
exists a $\sigma$ such that
$$
(\sigma, 0) \models \left[\bigwedge_h \always R_h\right] \land \always
(x \imp \always A). 
$$
By applying the $\tau_1$ transformation we have
$$
\tau_1[\always(x \imp \always A)] \longrightarrow \tau_1[\always(x \imp
\always {\bf y})] \land \tau_1[\always({\bf y} \imp A)]
$$
where ${\bf y}$ is a new proposition symbol.
Thus, we must show that there exists a model $\sigma'$ such that 
$$
(\sigma', 0) \models \left[\bigwedge_h \always R_h\right] \land
\always(x \imp \always {\bf y}) \land \always({\bf y} \imp A).
$$
First assume that $x$ is never satisfied in $\sigma$. 
A model $\sigma'$ identical to $\sigma$ except it contains the variable ${\bf y}$
such that ${\bf y}$ is false everywhere will suffice.
Otherwise let $j$ be the first place that $x$ is satisfied in $\sigma$. As $
(\sigma, 0) \models \always (x \imp \always A) $ for all $i \geq j$
then $(\sigma,i) \models A$. Let $\sigma'$ be the same as $\sigma$
except it contains a new proposition symbol ${\bf y}$ that is satisfied in all $i
\geq j$ and unsatisfied elsewhere i.e.\ $0 \leq i < j$. Thus, as
$\sigma'$ is identical to 
$\sigma$, except for ${\bf y}$, we have $(\sigma',i) \models A$ for all $i
\geq j$ and from the definition of $\sigma'$ we have for all $i \geq
j$, $(\sigma',i) \models {\bf y}$ and, for all $i < j$, $(\sigma',i) \models
\neg {\bf y}$. Thus, from the semantics of PLTL, $(\sigma',0) \models
\always({\bf y} \imp A)$. Now, as $(\sigma',i) \models {\bf y}$ for all $i \geq j$
then $(\sigma',j) \models \always {\bf y}$ from the semantics of $\always$.
Also, as  $(\sigma',j) \models x$ and by assumption $j$ is the first place
$x$ is satisfied in $\sigma$ and therefore $\sigma'$,
$(\sigma', 0) \models \always (x \imp \always {\bf y})$. Further 
$$(\sigma', 0) \models \bigwedge_h \always R_h$$
as
$$(\sigma, 0) \models \bigwedge_h \always R_h$$
 from our choice of $\sigma'$. Hence
$$
(\sigma', 0) \models \left[\bigwedge_h \always R_h\right] \land
\always(x \imp \always {\bf y}) \land \always({\bf y} \imp A)
$$
as desired. The proof of other renaming operations are similar.

Finally we consider the removal of unwanted temporal operators. Again,
we let $W$ be $\always A$ but this time assume that $A$ is a
literal. Assume that there exists a $\sigma$ such that
$$
(\sigma, 0) \models \left[\bigwedge_h \always R_h\right] \land \always
(x \imp \always A).
$$
By applying the $\tau_1$ transformation we obtain
$$
\tau_1[\always(x \imp \always A)] \longrightarrow \tau_1[\always(x
\imp A)] \land \tau_1[\always(x \imp {\bf y})] \land \always({\bf y} \imp \next A)
\land \always({\bf y} \imp \next {\bf y})
$$
where ${\bf y}$ is a new proposition symbol.
Thus, we must show that there exists a model $\sigma'$ such that 
$$
(\sigma', 0) \models \left[\bigwedge_h \always R_h\right] \land
\always(x \imp A) \land \always(x \imp {\bf y}) \land \always({\bf y} \imp \next
A) \land \always({\bf y} \imp \next {\bf y}).
$$
First assume that $x$ is never satisfied in $\sigma$. 
Similarly to the above, a model $\sigma'$ identical to $\sigma$ except
containing the variable ${\bf y}$ such that ${\bf y}$ is false everywhere 
will suffice.
Otherwise let $j$ be the first place that $x$ is satisfied in $\sigma$.
% (note that $j=\omega$ if $x$ is never satisfied). 
Let $\sigma'$ be the model
that is identical to $\sigma$ except it contains the variable ${\bf y}$ such
that for all $i \geq j$, $(\sigma',i) \models {\bf y}$ and for all $0 \leq i < j$,
$(\sigma',i) \models \neg {\bf y}$. Thus, as $\sigma$ is the same as
$\sigma'$ except for the valuation of ${\bf y}$, and 
$$(\sigma, 0) \models \bigwedge_h \always R_h$$
then, we have 
$$(\sigma', 0)\models \bigwedge_h \always R_h.$$
We have assumed that $(\sigma, 0) \models \always (x \imp \always A) $
so for all $i \geq j$, $(\sigma,i) \models A$ hence for all $i \geq j$,
$(\sigma',i) \models A$. Thus, as $(\sigma',j) \models x$, where $j$
is the first place that $x$ holds and for all $i \geq j$, $(\sigma',i)
\models A$ we have $(\sigma', 0) \models \always (x \imp A)$. Now as
$j$ is the first place that $x$ holds and $(\sigma',i) \models {\bf y}$ for
all $i \geq j$ we have $(\sigma',0) \models \always(x \imp {\bf y})$ and
$(\sigma',0) \models \always({\bf y} \imp \next {\bf y})$. Also, as $i \geq j$,
$(\sigma,i) \models A$ then, due to our choice of $\sigma'$, for all
$i \geq j$, $(\sigma',i) \models A$ and so $(\sigma',0) \models
\always ({\bf y} \imp \next A)$. Hence
$$
(\sigma', 0) \models \left[\bigwedge_h \always R_h\right] \land
\always(x \imp A) \land \always(x \imp {\bf y}) \land \always({\bf y} \imp \next
A) \land \always({\bf y} \imp \next {\bf y})
$$
as required.
\end{proof}

\begin{lemma}
%Let $\sigma$ be a model such that, for any PLTL formula $W$, then
%$(\sigma,0) \models W$. Then, there exists a model $\sigma'$ such that
%$(\sigma',0) \models \tau_0[W]$.
Given a model $\sigma$, and a PLTL formula $W$, such that $(\sigma,0) \models
W$, there exists a model $\sigma'$ such that $(\sigma',0) \models \tau_0[W]$.
\end{lemma}
\begin{proof}
Firstly note that if $(\sigma,0) \models W$ then there is a model
$\sigma''$ such that 
$$(\sigma'',0) \models (\lstart \imp {\bf y}) \land \always ({\bf y} \imp W).$$
The model $\sigma''$ is identical to $\sigma$ except it includes the
new proposition symbol ${\bf y}$ which is set to true where $i = 0$ and false
everywhere else. Applying $\tau_0$ to $W$, we obtain 
$$(\lstart \imp {\bf y}) \land \tau_1[\always({\bf y} \imp W)].$$
Now, from Lemma~\ref{lem.one.step}, and given that $(\lstart \imp {\bf y})
\land \always ({\bf y} \imp W)$ has a model $\sigma''$ every application of
the $\tau_1$ transformation can be satisfied in some new model. Hence,
if $W$ has a model then there exists a model that satisfies
$\tau_0[W]$.
\end{proof}
\begin{theorem}
A PLTL formula $A$ is satisfiable if, and only if, $\tau_0[A]$ is
satisfiable.
\end{theorem}
\begin{proof}
Lemmas 1 and 2 above show that if $\tau_0[A]$ is satisfiable in a
model, then $A$ is satisfiable in the same model. Lemmas 3 and 4 show
that, given a model for $A$, then we can construct a model for
$\tau_0[A]$. 
% The implication in the other direction is shown in \cite{Fisher97:JLC}. 
\end{proof}

\subsection{Example}
\label{eg.snf}
We illustrate the translation to the normal form by carrying out a
simple example transformation.
% In practice we use some shortcuts for example
Assume we want to show
$$
(\sometime p \land \always(p \imp \next p)) \imp \sometime \always p
$$
is valid. We negate, obtaining
$$
(\sometime p \land \always(p \imp \next p)) \land \always \sometime \neg p
$$
and begin to translate this into SNF.
First, we anchor to the beginning of time and split the conjuncts.
\begin{itemize}
\item[] $\begin{array}[t]{rrcl}
1. & \lstart & \imp & {\bf f} \\
2. & {\bf f} &  \imp & \sometime p \\
3. & {\bf f} & \imp & \always(p \imp \next p) \\
4. & {\bf f} & \imp & \always \sometime \neg p \\
\end{array}$
\end{itemize}
Formulae labelled 1 and 2 are now in normal form. We work on formula 3,
renaming the subformula $p \imp \next p$.
\begin{itemize}
\item[] $\begin{array}[t]{rrcl}
5. & {\bf f} & \imp & \always {\bf q} \\
6. & {\bf q} & \imp & (p \imp \next p) \\
\end{array}$
\end{itemize}
Next, we apply the $\always$ removal rules to formula 5 (to give 7, 8,
9 and 10) and rewrite formula 6 (to give 11).
\begin{itemize}
\item[] $\begin{array}[t]{rrcl}
7. & {\bf f} & \imp & {\bf q} \\
8. & {\bf f} & \imp & {\bf r} \\
9. & {\bf r} & \imp & \next {\bf q} \\
10. & {\bf r} & \imp & \next {\bf r} \\
11. & {\bf q} & \imp & (\neg p \lor \next p) \\
\end{array}$
\end{itemize}
Then, formulae 7 and 8 are rewritten into the normal form (giving 12-15)
and the subformula $\next p$ in formula 11 is renamed.
\begin{itemize}
\item[] $\begin{array}[t]{rrcl}
12. & \lstart & \imp & \neg {\bf f} \lor {\bf q} \\
13. & \ltrue & \imp & \next(\neg {\bf f} \lor {\bf q}) \\
14. & \lstart & \imp & \neg {\bf f} \lor {\bf r} \\
15. & \ltrue & \imp & \next(\neg {\bf f} \lor {\bf r}) \\
16. & {\bf q} & \imp & (\neg p \lor {\bf s}) \\
17. & {\bf s} & \imp & \next p \\
\end{array}$
\end{itemize}
Formula 16 is then rewritten into the correct form.
\begin{itemize}
\item[] $\begin{array}[t]{rrcl}
18. & \lstart & \imp & (\neg {\bf q} \lor \neg p \lor {\bf s}) \\
19. & \ltrue & \imp & \next (\neg {\bf q} \lor \neg p \lor {\bf s}) \\
\end{array}$
\end{itemize}
Next, we work on formula 4 renaming $\sometime \neg p$ with the new
proposition symbol $t$.
\begin{itemize}
\item[] $\begin{array}[t]{rrcl}
20. & {\bf f} & \imp  & \always {\bf t} \\
21. & {\bf t} & \imp & \sometime \neg p \\
\end{array}$
\end{itemize}
Then, we remove the $\always$ operator from formula 20 as previously
\begin{itemize}
\item[] $\begin{array}[t]{rrcl}
22. & {\bf f} & \imp &  {\bf t} \\
23. &   {\bf f} & \imp & {\bf u} \\
24. &    {\bf u} & \imp & \next {\bf t} \\
25. &    {\bf u} & \imp & \next {\bf u} \\
\end{array}$
\end{itemize}
and finally write formulae 22 and 23 into the normal form.
\begin{itemize}
\item[] $\begin{array}[t]{rrcl}
26. & \lstart & \imp &  \neg {\bf f} \lor {\bf t} \\
27. & \ltrue  & \imp &  \next (\neg {\bf f} \lor {\bf t}) \\
28. &   \lstart & \imp & \neg {\bf f} \lor {\bf u} \\
29. &   \ltrue & \imp & \next(\neg {\bf f} \lor {\bf u}) \\
\end{array}$
\end{itemize}
The resulting normal form is as follows.
\begin{itemize}
\item[] 
$\begin{array}{ll}
\begin{array}[t]{rrcl}
1. & \lstart & \imp & {\bf f} \\
2. & {\bf f} &  \imp & \sometime p \\
9. & {\bf r} & \imp & \next {\bf q} \\
10. & {\bf r} & \imp & \next {\bf r} \\
12. & \lstart & \imp & \neg {\bf f} \lor {\bf q} \\
13. & \ltrue & \imp & \next(\neg {\bf f} \lor {\bf q}) \\
14. & \lstart & \imp & \neg {\bf f} \lor {\bf r} \\
15. & \ltrue & \imp & \next(\neg {\bf f} \lor {\bf r}) \\
17. & {\bf s} & \imp & \next p \\
\end{array} 
\qquad \qquad \qquad \qquad \qquad
\begin{array}[t]{rrcl}
18. & \lstart & \imp & (\neg {\bf q} \lor \neg p \lor {\bf s}) \\
19. & \ltrue & \imp & \next (\neg {\bf q} \lor \neg p \lor {\bf s}) \\
21. & {\bf t} & \imp & \sometime \neg p \\
24. &    {\bf u} & \imp & \next {\bf t} \\
25. &    {\bf u} & \imp & \next {\bf u} \\
26. & \lstart & \imp &  \neg {\bf f} \lor {\bf t} \\
27. & \ltrue  & \imp &  \next (\neg {\bf f} \lor {\bf t}) \\
28. &   \lstart & \imp & \neg {\bf f} \lor {\bf u} \\
29. &   \ltrue & \imp & \next(\neg {\bf f} \lor {\bf u}) \\
\end{array}
\end{array}$
\end{itemize}

%%%%%%%%%%%%%%%%%%%%%%%%%%%% SECTION STARTS.............
\section{Resolution Rules}
\label{res.rules}
Once a formula has been transformed into SNF, both step resolution and
temporal resolution operations can be applied.
Step resolution effectively
consists of the application of the standard classical resolution rule
to formulae representing constraints at a particular moment in time,
together with simplification rules, subsumption rules, and rules for
transferring contradictions within states to constraints on previous
states. Temporal resolution resolves a sometime PLTL-clause whose right
hand side is, for example, $\sometime l$  with a set of SNF$_m$ PLTL-clauses
that together imply that $l$ is always false. We  also describe
{\em augmentation}, the addition of new variables required to translate the
resolvent from temporal resolution into SNF at the start of the
proof. This is useful in ensuring that no new proposition symbols need to
be added during the 
proof. 
\subsection{Step Resolution}
 Pairs of initial or step PLTL-clauses may be resolved using
the following (resolution) operations (where $A$ and $B$ are disjunctions
of literals, $C$ and $D$ are conjunctions of literals and $p$ is a
proposition). 
$$\begin{array}{rcl}
	      \lstart & \Rightarrow & A \lor p \\
	      \lstart  & \Rightarrow & B \lor \neg p \\ \hline
	      \lstart & \Rightarrow & A \lor B
\end{array}
\mbox{~~~~~~~~~~~}
\begin{array}{rcl}
	       C & \Rightarrow & \next(A \lor p) \\
	       D & \Rightarrow & \next(B \lor \neg p) \\ \hline
	       (C  \land D) & \Rightarrow & \next(A \lor B)
\end{array}$$
The following is used for PLTL-clauses which imply $\lfalse$ (where $A$ is
a conjunction of literals).
$$\begin{array}{rcl}
\{ A  \Rightarrow  \next \lfalse \} & \longrightarrow & 
\left \{
\begin{array}{rcl}
\lstart & \Rightarrow & \neg A \\
\ltrue & \Rightarrow & \next\neg A
\end{array}
\right \}
\end{array}$$
Thus, if, by satisfying $A$, a contradiction is produced in the next
moment, then $A$ must never be satisfied. The new constraints
generated effectively represent $\always\neg A$. This rewrite keeps
formulae in the suggested normal form and may, in turn, allow
further step resolution inferences to be carried out.

PLTL-clauses are kept in their simplest form by performing classical style
simplification, for example performing the following contraction
operations.
$$\begin{array}{rclcrcl}
 (l \land A \land l) & \Rightarrow & \next B & \longrightarrow & 
 (l \land A) & \Rightarrow & \next B \\
 (l \land A \land \neg l) & \Rightarrow & \next B & \longrightarrow & \lfalse &
 \Rightarrow & \next B \\ 
(A\land\ltrue)& \Rightarrow & \next B & \longrightarrow & A&
 \Rightarrow & \next B \\ 
(A\land\lfalse)& \Rightarrow & \next B & \longrightarrow & \lfalse &
 \Rightarrow & \next B \\ 
A  & \Rightarrow & \next (l \lor B \lor l) & \longrightarrow & A &
 \Rightarrow & \next (l \lor B) \\ 
A & \Rightarrow & \next (l \lor B \lor \neg l) & \longrightarrow & A &
 \Rightarrow & \next \ltrue \\
A & \Rightarrow & \next (B\llor\ltrue) & \longrightarrow & A &
 \Rightarrow & \next \ltrue \\ 
A & \Rightarrow & \next (B\llor\lfalse) & \longrightarrow & A &
 \Rightarrow & \next B
\end{array}$$
The following SNF PLTL-clauses can be removed during simplification as they
represent valid subformulae and therefore cannot contribute to the
generation of a contradiction.
$$\begin{array}{rcl}
\lfalse & \Rightarrow & \next A \\ 
A & \Rightarrow & \next \ltrue
\end{array}$$
The first PLTL-clause is valid as $\lfalse$ can never
be satisfied, and the second is valid as $\next \ltrue$ is always 
satisfied.

Subsumption also forms part of the step resolution process.  Here, as
in classical resolution, a PLTL-clause may be removed from the PLTL-clause-set if it
is subsumed by another PLTL-clause already present. Subsumption may be
expressed as the following operation.
$$\begin{array}{rcl}
\left\{
\begin{array}{rcl}
C & \Rightarrow & A \\
D & \Rightarrow & B
\end{array}
\right\}~~~
&
\stackrel{\vdash C \Rightarrow D ~~\vdash B\Rightarrow A} {\verylongrightarrow}
&
~~~~\{ D  \Rightarrow  B \}
\end{array}
$$
The side conditions $\vdash C \Rightarrow D$ and $\vdash B \Rightarrow
A$ must hold before this subsumption step can be applied and, in this
case, the PLTL-clause $C \imp A$ can be deleted without losing information.

The step resolution process terminates when either no new resolvents
can be generated or a contradiction is derived by generating % one of
the following unsatisfiable formula
$$
\lstart\ \Rightarrow\ \lfalse. % \qquad\quad \ltrue\ \Rightarrow\ \next\lfalse
$$

\subsection{Temporal Resolution}
\label{tresrule:old}
The temporal resolution operation effectively resolves together
formulae containing the `$\always$' and `$\sometime$'
connectives. However, the inductive interaction between the `$\next$'
and `$\!\always$' connectives in PLTL ensures that the application of
such an operation is non-trivial. Further, as the translation to SNF
restricts the PLTL-clauses to be of a certain form, the application of such
an operation will be between a sometime PLTL-clause and a {\em set\/} of step
PLTL-clauses that together ensure a complementary literal will {\em always}
hold.  Intuitively, temporal resolution may be applied between an
eventuality, i.e.\ a formula $\sometime l$ from the right-hand side of
a sometime PLTL-clause such as $C \Rightarrow \sometime l$, and a formula
which forces $l$ always to be false. Once the left-hand side of the
sometime PLTL-clause (i.e., $C$) is satisfied then, for the formula to be
satisfiable, there must be no other
PLTL-clauses forcing $l$ to always be false. To resolve with $C \Rightarrow
\sometime l$ then, a set of SNF$_m$ PLTL-clauses (see \S\ref{normal.form})
must be identified such that they characterise $A \imp \next \always
\neg l$ (where $A$ is in DNF)\footnote{The $\next$ operator occurs because
it is $\next \always \neg l$ rather than $\always \neg l$ that is actually
generated from a set of merged SNF step 
clauses.}.
So, the general temporal resolution
operation, written as an inference rule, becomes
$$
\begin{array}{rcl}
A  & \Rightarrow & \next \always \neg l \\
C  & \Rightarrow & \sometime l \\ \hline
C   & \Rightarrow & (\neg A) \unless l
\end{array}
$$
The intuition behind the resolvent is that, once $C$ has occurred then
$A$ must not be satisfied until $l$ has occurred (i.e.\ the
eventuality has been satisfied). 
% That is we can not allow $A$ to be satisfied
% once $C$ has been satisfied  as this would ensure that $l$ is false at all
% moments in the future.  
(Note that the generation of $C \imp (\neg A) \until l$ as a
resolvent would be sound. However as $(\neg A) \until l \equiv ((\neg
A) \unless l) \land \sometime l$ the resolvent would be equivalent to the
pair of resolvents $C \imp (\neg A) \unless l$ and  $C \imp \sometime l$.
The latter is subsumed by the sometime PLTL-clause we have resolved with.
So this leaves only the `$\unless$' formula.) The resolvent must next be
translated into SNF. 
In previous presentations, for example,~\cite{Fis90-resolve}, two
resolvents have been given. As the resolvent given here is sufficient
for completeness we omit the second.

In SNF we have no PLTL-clauses of the form $A \imp \next \always l$. So the
full temporal resolution operation applies between a sometime PLTL-clause and
a set of SNF$_m$ PLTL-clauses that together imply $A \imp \next \always \neg
l$. The temporal resolution operation, in detail, is
$$
\begin{array}{rcl}
A_0  & \Rightarrow & \next B_0\\
\ldots & \Rightarrow & \ldots\\
A_n  & \Rightarrow & \next B_n\\
C  & \Rightarrow & \sometime l \\ \hline
C   & \Rightarrow & \left[\displaystyle\bigwedge_{i=0}^n (\neg A_i)\right] \unless l
\end{array}
$$
with the side conditions that, for all $i$ $0 \leq i \leq n$,
$$
\begin{array}{l}
\prv B_i \Rightarrow \neg l; \mbox{ and}\\ 
\prv B_i \Rightarrow \displaystyle \bigvee_{j=0}^n A_j.
% \mbox{ for all } i, 0 \leq i \leq n.\\
\end{array}
$$
Here, the side conditions are simply propositional formulae so they must hold in
(classical) propositional logic. The first side condition ensures that by
satisfying any $B_i$ then $\neg l$ will be satisfied. The second shows that
once some $B_i$ is satisfied then one of the left hand sides ($A_j$) will
also be satisfied. Hence, if any $A_i$ is satisfied then, in the next moment,
$B_i$ is satisfied as is $\neg l$ as is $A_j$ for some $j$ and so on, so
that 
$$
(\bigvee_i A_i) \imp \next \always \neg l.
$$
The set of SNF$_m$ PLTL-clauses $A_i \imp \next B_i$ that satisfy these side
conditions are together known as {\em a loop in } $\neg l$. The disjunction
of the left hand side of this set of SNF$_m$ PLTL-clauses, i.e.\
$$
\bigvee_i A_i
$$
is known as a {\em loop formula} for $\neg l$.
The most complex part of this approach is the search for the set of
SNF$_m$ PLTL-clauses to use in the application of the temporal resolution
operation. Detailed explanation of the techniques developed for this
search is beyond the scope of this paper but is discussed at length
in~\cite{DFJ95-short,Dix96-CADE,Dix98:bfs}.

The resolvent must be translated into SNF before any further resolution
steps. A translation to the normal form is given below that avoids the 
renaming of the subformula 
$$
\bigwedge_{i=0}^n \neg A_i
$$
where $t$ is a new proposition symbol and $i = 0, \ldots, n$. Thus, for
each of the PLTL-clauses (\ref{res1}), (\ref{res2}) and (\ref{res5}) there
are $n+1$ copies, one for each $A_i$. (N.B., we 
will see in \S\ref{complete.sect} that this is important for completeness.)
% $$
\begin{eqnarray} %{rcl}
\lstart & \imp & \neg C \lor l \lor \neg A_i \label{res1} \\ 
\ltrue & \imp & \next( \neg C \lor l \lor \neg A_i) \label{res2} \\ 
\lstart & \imp & \neg C \lor l \lor t \label{res3} \\ 
\ltrue & \imp & \next( \neg C \lor l \lor t) \label{res4} \\
t & \imp & \next(l \lor \neg A_i) \label{res5} \\ 
t & \imp & \next(l \lor t)  \label{res6} 
\end{eqnarray}
%$$
%
We note that only the resolvents (\ref{res1}), (\ref{res2}) and
(\ref{res5}) depend on the particular loop being resolved with, i.e.\
contain a reference to $A_i$.

\subsection{Augmentation}
\label{augment}
The introduction of new variables,
such as $t$ above, makes proofs about the temporal resolution method more
difficult. Furthermore, if a temporal resolution
proof involves two temporal resolution inferences involving the same literal,
we may introduce two new variables where one would suffice. 
% To aid the
% proof of correctness we will use a modified resolution procedure which
Thus, for $n$ different eventualities we only require $n$ new proposition symbols.
We introduce these new proposition symbols at the start of the proof by adding 
the resolvents that do not contain $\neg A_i$, that is, have no reference
to the loop detected  (i.e.\ the PLTL-clauses above labelled \ref{res3},
\ref{res4} and \ref{res6}) at the beginning and the rest of the
PLTL-clauses, if required, as the proof proceeds. The following definitions formalise
this technique. Given an eventuality $\sometime l$, the new proposition symbol
introduced is $w_l$ (rather than $t$ above) which can be thought of as {\em
waiting for} $l$. Hence having translated to SNF and augmented, we can be
sure that no new proposition symbols appear during the application of the
resolution rules. 

\begin{defn}{[Augmented PLTL-Clause Sets]} 
Given a set, $S$, of SNF PLTL-clauses, we construct an augmented set of PLTL-clauses
$Aug(S)$ as follows. For each literal $l$ which occurs as an
eventuality in $S$ we introduce a new proposition symbol, $w_l$, and record
the correspondence between $l$ and $w_l$. The variable $w_l$ will be
used to record the condition that we are {\em waiting\/} for $l$ to
occur. The first defining PLTL-clause for $w_l$ is
\begin{equation}
w_l \imp \next (l \lor w_l). \label{wl1}
\end{equation}
Then, for each PLTL-clause $C \imp \sometime l$, we add both
% $$
\begin{eqnarray} 
\lstart & \imp & \neg C \lor l \lor w_l \label{wl2} \\
\ltrue  & \imp & \next(\neg C \lor l \lor w_l) \label{wl3}.
\end{eqnarray}
% $$
%
\end{defn}

\begin{defn}
The {\em loop resolvents} for a sometime PLTL-clause $C \imp \sometime l$ and a
loop formula $\bigvee_i A_i$ are
% $$
\begin{eqnarray}
\lstart & \imp & \neg C \lor l \lor \neg A_i  \label{wl4}\\ 
\ltrue & \imp & \next( \neg C \lor l \lor \neg A_i) \label{wl5}\\ 
w_l & \imp & \next(l \lor \neg A_i) \label{wl6} 
\end{eqnarray}
% $$
for each $i$.
\end{defn}
Note, the loop resolvents for a particular sometime clause and loop formula
are the only clauses added to the clause-set by applying the temporal
resolution rule. 

\subsection{An Algorithm for the Temporal Resolution Method}
\label{alg.full.tr}
Given any temporal formula, $A$, to be tested for unsatisfiability,
the following steps are performed.
\begin{enumerate}
\item Translate $A$ into SNF, giving $A_s$.
\item Augment $A_s$, giving $Aug(A_s)$.
\item Perform step resolution (including simplification and subsumption)
      on  $Aug(A_s)$ until either 
 \begin{enumerate}
 \item $\lstart \imp \lfalse$ is derived --- terminate noting that $A$ is unsatisfiable; or 
 \item no new resolvents are generated --- continue to step (4).
 \end{enumerate}
\item Select an eventuality from the right-hand side of a sometime PLTL-clause
 within $Aug(A_s)$, for example $\sometime l$. Search for loop-formulae for
 $\neg l$.
\item 
Construct loop resolvents for the loop-formulae detected and each sometime
PLTL-clause with $\sometime l$ on the right-hand side.
If any new formulae
 (i.e.\ that are not subsumed by PLTL-clauses already present) have been
 generated, go to step~(3). 
\item If all eventualities have been resolved, terminate declaring $A$
satisfiable, otherwise go to step~(4).
\end{enumerate}
We will consider the soundness, completeness and termination of this method in
\S\ref{sec.correct}. 
%
%%%%%%%%%%%%%%%%%%%%%%%%%%%% SECTION STARTS.............
\section{Examples}
\label{sec.eg}
We illustrate the method by presenting a selection of examples.
\subsection{Step Resolution Example}
% example with only step/initial resolution --- to show link to
% classical resolution
We prove an instance of one of the PLTL axioms that requires only step
resolution, namely 
$$
\vdash\ \next (a \imp b) \imp (  \next a \imp \next b).
$$
We negate
$$
\next (a \imp b) \land (\next a \land \next \neg b)
$$
and rewrite into SNF as follows.
\begin{itemize}
\item[] $\begin{array}[t]{rrcl}
1. & \lstart & \imp & f \\
2. & f & \imp & \next x \\
3. & \lstart & \imp &  (\neg x \lor \neg a \lor b) \\
4. & \ltrue & \imp & \next (\neg x \lor \neg a \lor b) \\
5. & f & \imp & \next a \\
~6. & f & \imp & \next \neg b 
\end{array}$
\end{itemize}
There are no sometime PLTL-clauses so augmentation adds no new PLTL-clauses.
Resolution can be carried out as follows.
\begin{itemize}
\item[] $\begin{array}[t]{rrcll}
7. & f & \imp & \next (\neg x \lor \neg a)\qquad & [4,6 \mbox{ Step Resolution}] \\
8. & f  & \imp & \next \neg x  & [5,7 \mbox{ Step Resolution}] \\
9. & f  & \imp & \next \lfalse & [2,8 \mbox{ Step Resolution}] \\
10. & \lstart  & \imp & \neg f & [9 \mbox{ Rewriting}] \\
11. & \ltrue  & \imp & \next \neg f & [9 \mbox{ Rewriting}] \\
12. & \lstart  & \imp & \lfalse & [1,10 \mbox{ (Initial) Step Resolution}]
\end{array}$
\end{itemize}
A contradiction has been obtained meaning the negated formula is
unsatisfiable and therefore the original formula is valid.
\subsection{Temporal Resolution Example (From a Set of Clauses)}
\label{eg.tr}
Assume we wish to show that the following set of PLTL-clauses (already translated
into SNF) is unsatisfiable.
\begin{itemize}
\item[] $\begin{array}[t]{rrcl}
1. & \lstart & \imp & f \\
2. & \lstart & \imp & a \\
3. & \lstart & \imp & p \\
4. & f & \imp & \sometime \neg p \\
5. & f & \imp & \next a \\
6. & a & \imp & \next (b \lor x) \\
7. & b & \imp & \next a \\
8. & b & \imp & \next p \\
9. & a & \imp & \next p \\
10. & a & \imp & \next \neg x
\end{array}$
\end{itemize}
As the set of PLTL-clauses contains a sometime PLTL-clause (no. 4) we
augment with the following PLTL-clauses.
\begin{itemize}
\item[]$\begin{array}[t]{rrcll}
11.& \lstart & \imp & \neg f \lor \neg p \lor w_{\neg p} &
[4 \mbox{ Augmentation}] \\
12. & \ltrue & \imp & \next( \neg f \lor \neg p \lor w_{\neg p}) &
[4 \mbox{ Augmentation}]   \\
13. & w_{\neg p} & \imp & \next(\neg p \lor w_{\neg p}) &
[4 \mbox{ Augmentation}]   \\
\end{array}$
\end{itemize}
Step resolution occurs as follows.
\begin{itemize}
\item[] $\begin{array}[t]{rrcll}
14.& a & \imp & \next b \qquad & [6,10 \mbox{ Step Resolution}] 
\end{array}$
\end{itemize}
Note other step resolution inferences may be performed, for example between
1 and 11 but we omit them as they play no part in the proof.
By merging PLTL-clauses 9 and 14, and 7 and 8 into SNF$_m$ using the
merged-SNF rule given in \S\ref{sub.snf} we obtain the
following loop in 
$p$ (in SNF${}_m$)
$$
\begin{array}{rcll}
a & \imp & \next (b \land p)\quad \qquad & [9,14 \mbox{ SNF}_m]\\
b & \imp & \next (a \land p) & [7,8 \mbox{ SNF}_m]
\end{array}
$$
for resolution with PLTL-clause 4. 
The resolvents after temporal resolution are PLTL-clauses 15--20 below
\begin{itemize}
\item[] $\begin{array}[t]{rrcll}
15. & \lstart & \imp & \neg f \lor \neg p \lor \neg a \qquad & [4,7,8,9,14 \mbox{ Temporal Resolution}]\\ 
16. & \ltrue & \imp & \next( \neg f \lor \neg p \lor \neg a) \qquad & [4,7,8,9,14 \mbox{ Temporal Resolution}]\\
17. & \lstart & \imp & \neg f \lor \neg p \lor \neg b \qquad & [4,7,8,9,14 \mbox{ Temporal Resolution}]\\
18. & \ltrue & \imp & \next( \neg f \lor \neg p \lor \neg b) \qquad & [4,7,8,9,14 \mbox{ Temporal Resolution}]\\
19. & w_{\neg p} & \imp & \next(\neg p \lor \neg a) \qquad & [4,7,8,9,14 \mbox{ Temporal Resolution}]\\
20. & w_{\neg p} & \imp & \next(\neg p \lor \neg b)  \qquad & [4,7,8,9,14 \mbox{ Temporal Resolution}]\\
\end{array}$
\end{itemize}
and the proof concludes as follows.
\begin{itemize}
\item[] $\begin{array}[t]{rrcll}
21. & \lstart & \imp & \neg f \lor \neg a  & [3,15 \mbox{ (Initial) Step Resolution}]\\
22. & \lstart & \imp & \neg f  & [2,21 \mbox{ (Initial) Step Resolution}]\\
23. & \lstart & \imp & \lfalse & [1,22 \mbox{ (Initial) Step Resolution}] \\
\end{array}$
\end{itemize}
A contradiction has been obtained hence the set of PLTL-clauses is unsatisfiable.
\subsection{Temporal Resolution Example (From a Formula)}
Next we show that $\always a \land \sometime \neg a$ is
unsatisfiable. First we translate to the normal form.
\begin{itemize}
\item[] $\begin{array}[t]{rrcl}
1. & \lstart & \imp & x \\
2. & x & \imp & \sometime \neg a \\
3. & \lstart & \imp & \neg x \lor a \\
4. & \ltrue  & \imp & \next (\neg x \lor a) \\
5. & \lstart & \imp & \neg x \lor y \\
6. & \ltrue  & \imp & \next (\neg x \lor y) \\
7. & y & \imp & \next y \\
8. & y & \imp & \next a \\
\end{array}$
\end{itemize}
As the set of PLTL-clauses contains a sometime PLTL-clause (no. 2) we
augment with the following PLTL-clauses.
\begin{itemize}
\item[]$\begin{array}[t]{rrcll}
9.& \lstart & \imp & \neg x \lor \neg a \lor w_{\neg a} &
[2 \mbox{ Augmentation}] \\
10. & \ltrue & \imp & \next( \neg x \lor \neg a \lor w_{\neg a}) &
[2 \mbox{ Augmentation}]   \\
11. & w_{\neg a} & \imp & \next(\neg a \lor w_{\neg a}) &
[2 \mbox{ Augmentation}]   \\
\end{array}$
\end{itemize}
We can find a loop for resolution with PLTL-clause 2 by merging 7 and 8 to
give
$$ y \imp  \next (y \land a).$$
One of the resolvents obtained is PLTL-clause 12 from which we can derive a
contradiction. 
\begin{itemize}
\item[] $\begin{array}[t]{rrcll}
12. & \lstart & \imp & \neg x \lor \neg a \lor \neg y \qquad & [2,7,8 \mbox{ Temporal Resolution}]\\ 
13. & \lstart & \imp & \neg x \lor \neg a \qquad & [5,12 \mbox{ (Initial) Step Resolution}]\\ 
14. & \lstart & \imp & \neg x \qquad & [3,13 \mbox{ (Initial) Step Resolution}]\\ 
15. & \lstart & \imp & \lfalse \qquad & [1,14 \mbox{ (Initial) Step Resolution}]\\ 
\end{array}$
\end{itemize}
\subsection{A Larger Example}
% a `big' example --- showing a more realistic scenario
Here we conclude the example introduced in \S\ref{eg.snf}. Recall we are
trying to show that 
$$
(\sometime p \land \always(p \imp \next p)) \imp \sometime \always p
$$
is valid. We negated and translated the formula into SNF in
\S\ref{eg.snf}. The PLTL-clauses in normal
form are repeated here although they have been renumbered sequentially. We
only show the steps relevant to the refutation.
\begin{itemize}
\item[] $\begin{array}{ll}
\begin{array}[t]{rrcl}
1. & \lstart & \imp & f \\
2. & f &  \imp & \sometime p \\
3. & r & \imp & \next q \\
4. & r & \imp & \next r \\
5. & \lstart & \imp & \neg f \lor q \\
6. & \ltrue & \imp & \neg f \lor q \\
7. & \lstart & \imp & \neg f \lor r \\
8. & \ltrue & \imp & \neg f \lor r \\
9. & s & \imp & \next p \\
\end{array} \qquad \qquad \qquad \qquad \qquad
\begin{array}[t]{rrcl}
10. & \lstart & \imp & (\neg q \lor \neg p \lor s) \\
11. & \ltrue & \imp & \next (\neg q \lor \neg p \lor s) \\
12. & t & \imp & \sometime \neg p \\
13. &    u & \imp & \next t \\
14. &    u & \imp & \next u \\
15. & \lstart & \imp &  \neg f \lor t \\
16. & \ltrue  & \imp &  \next (\neg f \lor t) \\
17. &   \lstart & \imp & \neg f \lor u \\
18. &   \ltrue & \imp & \next(\neg f \lor u) \\
\end{array}
\end{array}$
\end{itemize}
Next we augment the set of PLTL-clauses to account for the two sometime
PLTL-clauses 2 and 12.
\begin{itemize}
\item[] $\begin{array}[t]{rrcll}
19. & \lstart & \imp & ( \neg f \lor w_{p} \lor p) &
[2 \mbox{ Augmentation}] \\
20. & \ltrue & \imp & \next( \neg f \lor w_{p} \lor p) &
[2 \mbox{ Augmentation}] \\
21. & w_{p} & \imp & \next(w_{p} \lor p) &
[2 \mbox{ Augmentation}] \\
22. & \lstart & \imp & ( \neg t \lor w_{\neg p} \lor \neg p) &
[12 \mbox{ Augmentation}] \\
23. & \ltrue & \imp & \next( \neg t \lor w_{\neg p} \lor \neg p) &
[12 \mbox{ Augmentation}] \\
24. & w_{\neg p} & \imp & \next(w_{\neg p} \lor \neg p) &
[12 \mbox{ Augmentation}] \\
\end{array}$
\end{itemize}
Step resolution then begins.
\begin{itemize}
\item[] $\begin{array}[t]{rrcll}
25. & r & \imp & \next(\neg p \lor s)\qquad & [3,11 \mbox{ Step Resolution}] \\
26. & (s \land r) & \imp & \next s & [9,25 \mbox{ Step Resolution}] \\
\end{array}$
\end{itemize}
By merging PLTL-clauses 4, 9 and 26 into SNF$_m$ we obtain the loop
$$
(s \land r) \imp \next (s \land r \land p)
$$
for resolution with PLTL-clause 12. 
% This generates the resolvent
% $$
% t \imp \neg(s \land r) \unless \neg p
% $$
% which can be translated into SNF as described in Section~\ref{tresrule:old}
This generates additional PLTL-clauses (from the resolvent) as follows.
% where $w_{\neg p}$ is a new proposition. 
\begin{itemize}
\item[] $\begin{array}[t]{rrcll}
27. & \lstart & \imp & ( \neg t \lor \neg s \lor \neg r \lor \neg p) &
[4,9,26,12 \mbox{ Temporal Resolution}] \\
28. & \ltrue & \imp & \next( \neg t \lor \neg s \lor \neg r \lor \neg p) &
[4,9,26,12 \mbox{ Temporal Resolution}] \\
29. & w_{\neg p} & \imp & \next( \neg s \lor \neg r \lor \neg p) &
[4,9,26,12 \mbox{ Temporal Resolution}] \\
\end{array}$
\end{itemize}
Thus the refutation continues as follows.
\begin{itemize}
\item[] $\begin{array}[t]{rrcll}
30. & \ltrue & \imp & \next( \neg t \lor \neg r \lor \neg p \lor \neg q)\quad &
[11,28 \mbox{ Step Resolution}] \\
31. & r & \imp & \next( \neg t  \lor \neg p \lor \neg q) &
[4,30 \mbox{ Step Resolution}] \\
32. & r & \imp & \next( \neg t  \lor \neg p) & [3,31 \mbox{ Step Resolution}] \\
33. & (r \land u) & \imp & \next \neg p & [13,32 \mbox{ Step Resolution}] \\
\end{array}$
\end{itemize}
Now by merging PLTL-clauses 4, 14 and 33
$$
(r \land u) \imp \next (r \land u \land \neg p)
$$
we have a loop for resolution with PLTL-clause 2, which generates 
% the resolvent
% $$
% f \imp \neg(r \land u) \unless p
% $$
several resolvents, including PLTL-clause 34.
\begin{itemize}
\item[] $\begin{array}[t]{rrcll}
34. & \lstart & \imp & (\neg f \lor \neg r \lor \neg u \lor p) & [2,4,14,33 \mbox{ Temporal Resolution}]  \\
35. & \lstart & \imp & (\neg f \lor \neg r \lor \neg u \lor \neg q \lor s) & [10,34 \mbox{ (Initial) Step Resolution}]  \\
36. & \lstart & \imp & (\neg f \lor \neg r \lor \neg u \lor \neg q \lor
\neg t \lor \neg p)\quad & [27,35 \mbox{ (Initial) Step Resolution}]  \\
37. & \lstart & \imp & (\neg f \lor \neg r \lor \neg u \lor \neg q \lor
\neg t) & [34,36 \mbox{ (Initial) Step Resolution}]  \\
38. & \lstart & \imp & (\neg f \lor \neg r \lor \neg q \lor \neg t) & [17,37 \mbox{ (Initial) Step Resolution}]  \\
39. & \lstart & \imp & (\neg f \lor \neg r \lor \neg q) & [15,38 \mbox{
(Initial) Step Resolution}]  \\
40. & \lstart & \imp & (\neg f \lor \neg q) & [7,39 \mbox{ (Initial) Step Resolution}]  \\
41. & \lstart & \imp & \neg f  & [5,40 \mbox{ (Initial) Step Resolution}]  \\
42. & \lstart & \imp & \lfalse  & [1,41 \mbox{ (Initial) Step Resolution}]  \\
\end{array}$
\end{itemize}

%%%%%%%%%%%%%%%%%%%%%%%%%%%% SECTION STARTS.............
\section{Correctness}
\label{sec.correct}
First we show that augmentation is satisfiability preserving. Next, a
soundness result is obtained by showing that an application of the step or
temporal resolution rule preserves satisfiability. Finally completeness is
proved by considering the construction of a graph representing all possible
models of the augmented set of PLTL-clauses. Here, deletions of parts of the
graph that cannot be used to construct models are associated with step and
resolution rules. 

\subsection{Augmented PLTL-Clause Sets}
\label{mod.rulesets}

\noindent We will show that an augmented PLTL-clause set has a {\em
model\/} if, and only if, its underlying (non-augmented) PLTL-clause set has
a model. 

\begin{defn}
Given a set, $S$, of SNF PLTL-clauses, a {\em
normal model} for the augmented PLTL-clause set for $S$ is a model which
satisfies the formula
\end{defn}
\begin{equation}
\always (w_l \iff  (\neg l \land \sometime l)) \label{wl7}
\end{equation}
for each literal $l$ which occurs as an eventuality (i.e.\ inside the scope
of a $\sometime$ operator) in $S$. 

\begin{defn} An augmented PLTL-clause set is said to be {\em well-behaved\/}
if it is either unsatisfiable or has a normal model.
\end{defn}

\begin{lemma}[Augmentation] 
\label{wellbehaved}
If $S$ is a set of SNF PLTL-clauses then
\begin{enumerate}
\item $Aug(S)$ is well-behaved, and,
\item $Aug(S)$ has a model if and only if $S$ has a model.
\end{enumerate}
\end{lemma}
\begin{proof}
If $Aug(S)$ has a model then, ignoring the value of each $w_l$ at each
moment gives a model for $S$. Conversely, if $S$ has a model $M$, then
$M$ can be extended to a model $M'$ for $Aug(S)$ by giving $w_l$ the
same truth value as $\neg l \land \sometime l$ in $M$ in each state, and for each
literal $l$. The model $M'$ clearly satisfies the formulae (\ref{wl1}),
(\ref{wl2}) and (\ref{wl3}) from \S\ref{augment} and (\ref{wl7}) above. The
lemma follows easily from these two observations.
\end{proof}

\vspace{1ex}

\subsection{Soundness}
\label{sound.sect}

\subsubsection{Step Resolution Rules}
It is easy to see that given a satisfiable set of PLTL-clauses the application of
the initial or step resolution inferences, or simplification preserves
satisfiability. 
\subsubsection{Temporal Resolution Rule}
The following lemma is a soundness result for the temporal resolution rule
(applied to augmented PLTL-clause sets).

\begin{lemma}[Soundness]
\label{sound}
Let $S$ be a well-behaved augmented PLTL-clause set. Let the PLTL-clause set $T$ be
obtained from $S$ by application of the temporal resolution operation. Then
\begin{enumerate}
\item $T$ is well-behaved, and,
\item if $S$ is satisfiable then $T$ is satisfiable.
\end{enumerate}
\end{lemma}

\begin{proof}
If $S$ is satisfiable then $S$ has a model, and by Lemma~\ref{wellbehaved}
it has a normal model $M$. The side 
conditions for temporal resolution guarantee that the loop resolvents i.e.\
formulae
(\ref{wl4}), (\ref{wl5}) and (\ref{wl6}) given in \S\ref{augment} hold
in $M$, and so $M$ 
is a (normal) model for $T$, i.e.\ $T$ is satisfiable.  
If $S$ is unsatisfiable then the addition of PLTL-clauses to produce $T$ is
also unsatisfiable. Hence $T$ is well-behaved.
\end{proof}

\subsection{Completeness}
\label{complete.sect}
We will now prove the completeness of the temporal resolution
procedure by induction on the size of a {\em behaviour graph} of a set
of SNF PLTL-clauses. Note, as we have added all the new variables
required for the translation of the unless operator by augmentation in
\S\ref{mod.rulesets} and avoided renaming the conjunction that
occurs from negating the loop-formula (a disjunction) as mentioned in
\S\ref{tresrule:old} we require no new proposition symbols during the
proof. Thus the graph constructed has all the propositional symbols we
require and will not increase in size during the proof.

\begin{defn}{[Behaviour Graph]}
Given a set $S$ of SNF PLTL-clauses, we construct a finite directed graph $G$
as follows. The nodes of $G$ are all ordered pairs $(V,E)$ where $V$
is a valuation of the proposition symbols occurring in $S$ and $E$ is a
subset of the literals occurring as eventualities in $S$ i.e.\ literals
occurring on the right-hand side of the sometime PLTL-clauses in $S$. Thus
$V$ contains either $p$ or $\neg p$ for each proposition symbol $p$ in $S$. For each
node $(V,E)$, let $R$ be the set of step PLTL-clauses of $S$ which are
``fired'' by $V$ --- that is, the set of step PLTL-clauses whose left-hand
sides are satisfied by $V$. Let $L$ be the set of clauses on the
right-hand sides of the PLTL-clauses in $R$, i.e.\ $L$ contains
formulae that are the disjunction of literals from the  right-hand side of each PLTL-clause in
$R$ having first removed the next operator.  Let $E'$ be the set of elements of
$E$ which are not satisfied by $V$.  For each valuation $V'$ which
satisfies $L$, let $E''$ be the set of literals occurring on the
right-hand sides of the sometime PLTL-clauses fired by $V'$.  Then for
each $V'$ construct an edge in $G$ from $(V,E)$ to $(V',E' \cup E'')$.
These are the only edges originating from $(V,E)$. Let $L_0$ be
the set of initial PLTL-clauses of $S$. For each valuation $V$ which
satisfies $L_0$, where $E'$ is the set of literals occurring on the
right-hand sides of the sometime PLTL-clauses fired by $V$, the node
$(V,E')$ is designated as an {\em initial node} of $G$. The behaviour
graph of $S$ is the full subgraph of $G$ given by the set of nodes
reachable from the initial nodes. We regard the identification of the
initial nodes as part of the structure of the behaviour graph.
\end{defn}

\begin{lemma}
\label{add.rules}
Let $S$ be a set of SNF PLTL-clauses and let $T$ be the set of SNF PLTL-clauses
obtained from $S$ by adding finitely many initial PLTL-clauses and finitely
many step PLTL-clauses which only involve proposition symbols occurring in $S$. Then
the behaviour graph of $T$ is a subgraph of the behaviour graph of
$S$.
\end{lemma}

\begin{proof}
This is established by
induction on the length of the shortest path from an initial node to 
an arbitrary node in the behaviour graph of $T$. Let $len$ be the length of the
shortest path from an  initial node to a node $n$. To show the base case we
let $len =0$ and show that any initial node in the behaviour
graph of $T$ is an initial node in the behaviour graph of $S$. 
Let $I \subseteq S$ be the initial PLTL-clauses of $S$ and $I' \subseteq T$ the
initial PLTL-clauses of $T$. As $T$ has been constructed by adding initial and/or
step  PLTL-clauses to $S$, $I \subseteq I'$. Take any initial node
$n_0 = (V_0,E_0)$ in the behaviour graph for $T$. From the
definition of the 
behaviour graph $V_0$ must satisfy the right hand side of the PLTL-clauses in
$I'$. As $I \subseteq I'$ then $V_0$ must also satisfy the 
right hand side of the PLTL-clauses in $I$. As the set of sometime PLTL-clauses in
$S$ and $T$ are unchanged, i.e.\ as $V_0$ satisfies the left hand side of the
same sometime PLTL-clauses in $S$ and $T$ the set $E_0$ will be the same in each
graph for $V_0$ and thus the node $n_0 = (V_0,E_0)$ is also in the
behaviour graph for $S$.

Next we assume that if any node $n$,  where the length of the shortest
path from an initial node to $n$ is $m$, is
in the behaviour  graph for $T$, it is also in the behaviour graph for $S$.
We show that any node $n'$ in the behaviour  graph for $T$ whose shortest
path length from an initial node is $m+1$,  is also in the behaviour graph for
$S$. 
%
% prove that any non-initial node in the behaviour graph of $T$ is
% present in the behaviour graph of $S$. 
Let $J \subseteq S$ be the step PLTL-clauses in $S$ and $J' \subseteq T$ the
step PLTL-clauses in $T$. By assumption we have $J \subseteq J'$.
Consider some node $n' = (V',E')$ in the behaviour graph of $T$ where the shortest
path from an initial node to $n'$ is $m+1$. Let $n = (V,E)$ be any node in
the behaviour graph for $T$ such that there is an edge from $n$ to $n'$ and
the shortest path from an initial node to $n$ is of length $m$.
%
% that is reached
% by following $1$ edge from some node $n = (V,E)$ which is $len = m$ from
% some initial node. 
By the induction hypothesis, we assume that $n$ is also in the
behaviour graph for $S$.

Let $X' \subseteq J'$ be the set of step PLTL-clauses in $T$ such that  the left
hand sides are satisfied by $V$ and the right hand side satisfy $V'$.
Let $X \subseteq J$ be the corresponding set of step PLTL-clauses in $S$
i.e.\ where the left hand sides are satisfied by $V$ and the right hand side
satisfy $V'$ . 
% Thus as $J \subseteq J'$ also satisfy $X \subseteq X'$ the relevant set
% of formula for the behaviour graph of $S$. 
As $J \subseteq J'$  we have $X \subseteq X'$.
Furthermore as no change has
been made to the set of sometime PLTL-clauses any eventualities outstanding
from $n$ or triggered by $n'$ will be the same in each graph. Thus $n'$ is
also present in the behaviour graph for $S$.
% Thus any edge in the behaviour graph for $T$ must also be present in the
% graph for $S$.
\end{proof}
\vspace{1ex}
\begin{lemma}
\label{any.model}
Any model for a set of SNF-PLTL-clauses, $S$, can be constructed from a path
through the behaviour graph for $S$.
\end{lemma}
\begin{proof}
To construct a model from a suitable path, $N_0,N_1,N_2,\ldots$ where each
$N_i = (V_i,E_i)$, through 
the behaviour graph (i.e.\ one which is infinite and all
eventualities are satisfied) take the valuation $V_i$ from
each node $N_i$ in the path (and delete any negated proposition symbols). Any
proposition symbols that do not occur in $S$ but 
are required in the model may be set arbitrarily. Details of how to
construct models from behaviour graphs are given in
Lemma~\ref{empty.unsat}.

Take any model $\sigma = s_0, s_1, \ldots$ for $S$. We show that
this model can be constructed from a path through the behaviour
graph. First delete any 
proposition symbols not in $S$ from $\sigma$ to give $\sigma' = s'_0, s'_1,
\ldots$. As these proposition symbols do not occur in $S$ they have no constraints
on them so by setting these proposition symbols to true and false in the correct
way we can 
recover $\sigma$. Note that $\sigma'$ is a model for $S$. By definition the
behaviour graph for $S$ is the reachable subgraph from the set of 
initial nodes. The behaviour graph has been constructed where the $V$
component of each node consists of every possible valuation. Let ${\rm
pos}(V_i)$ be the set of non-negated proposition symbols in $V_i$. As $\sigma'$
is a model for $S$, $s'_0$ must satisfy the initial rules $I \subseteq S$. To
construct the behaviour graph for $S$ the initial nodes are those with
valuations that satisfy $I$, for a particular $E$ component. As nodes are
constructed with each valuation and subset of eventualities there must be a
node $N_0 =(V_0,E_0)$ where ${\rm pos}(V_0) = s'_0$.

Next for some $s'_i$ in $\sigma'$ assume that there is a node $N_i
=(V_i,E_i)$ in the behaviour graph for $S$ such that ${\rm pos}(V_i) = s'_i$.
We show that ${\rm pos}(s'_{i+1}) = V_{i+1}$ for some node $N_{i+1}
=(V_{i+1},E_{i+1})$ in the behaviour graph for $S$. Let $R \subseteq S$ be
the set of step PLTL-clauses in $S$. Take the set of step
PLTL-clauses $R' 
\subseteq R$ such that the left hand side of the PLTL-clauses in $R'$ is
satisfied by $V_i$. As  ${\rm pos}(V_i) = s'_i$, $s'_i$ must satisfy the left hand
side of the PLTL-clauses in $R'$. As $\sigma'$ is a model for $S$,
$s'_{i+1}$ must satisfy the right hand side of each PLTL-clauses in $R'$
having deleted the next operator. From the construction of the behaviour
graph, edges are drawn from $N_i$ to nodes whose valuation satisfies 
the right hand side of each PLTL-clauses in $R'$ having deleted the next
operator (for some $E$ component). As nodes have been constructed for all
valuation/eventuality component combinations there will be one
$N_{i+1}=(V_{i+1},E_{i+1})$ such that ${\rm pos}(V_{i+1})=s'_{i+1}$.

Hence we can construct $\sigma'$ using the valuations from each node and
following a path through the behaviour graph for $S$. This can be extended
to $\sigma$ by setting the additional proposition symbols as required.
\end{proof}

\begin{lemma}\label{simp.sump}
Let $S$ be a set of PLTL-clauses and $T$ be the set of clauses obtained
from $S$ by applying one simplification or subsumption step. The behaviour
graph for $S$ is the same as the behaviour graph for $T$.
\end{lemma}
\begin{proof}
%\begin{fbox}
%Fix me!
%\end{fbox}

First assume  we have performed a simplification step.
We show that any node and edge that is in the behaviour graph for $S$ is
also in the behaviour graph for $T$. The proof of the converse is similar.
The proof is by induction on the
length of the shortest path from an initial node. 
For the base case the length of the path from an initial node to $n$ is 0,
i.e.\ $n$ is an initial node. If the simplification step has not been
performed on an initial PLTL-clause i.e.\ the set of initial PLTL-clauses in
$S$ and in $T$ are the same then $n$ must also be in the behaviour graph for
$T$. Otherwise we have performed a simplification step on an initial PLTL-clause
i.e $S$ contains $\lstart \imp Y$ and $T$ contains $\lstart \imp Y'$ where
$Y \equiv Y'$. Each initial node $n$ in the behaviour graph for $S$ satisfies
$Y$ by definition of the behaviour graph. As $Y \equiv Y'$ node $n$ also
satisfies $Y'$ so $n$ is in the behaviour graph for $T$.

Next assume the node $n$ in the behaviour graph for $S$, whose shortest
 path distance from an initial  node is $m$, is also in the behaviour graph
 for $T$. We show that any node of shortest path length $m+1$ from an
 initial node is also in the behaviour graph for $T$. Take a node $n''$
 in the behaviour graph for $S$ whose shortest path length from an  initial
 node is $m+1$. Consider $n'$ such that $(n',n'')$ is an edge in the
 behaviour graph from $S$ where the shortest path length from $n'$ to an
 initial node is $m$. From the induction hypothesis $n'$ is also in also in
 the behaviour graph for $T$.
 Assume that a simplification step has been applied to rule $X \imp \next Y
\in S$ to obtain $X' \imp \next Y' \in T$ and that $n'$ satisfies $X$.
Thus from the definition of the behaviour graph $n''$ must satisfy $Y$.
 As we have performed a simplification step $X \equiv X'$ and $Y \equiv Y'$
 so $n'$ also satisfies $X'$ and $n''$ satisfies $Y'$ as the sets $S$ and
 $T$ are unchanged apart from this. Hence $n''$ and the edge $(n',n'')$ must
 also  be in $T$. If the node $n'$ didn't satisfy $X$, or the
 simplification rule had been on an initial PLTL-clause then $n''$ would
 again be in the behaviour graph for $T$ as the remaining rules are
 unchanged. The proof of the converse is similar.

To show the proof holds for a subsumption step assume $S$ contain rules $X
\imp \next Y$ and $X' \imp \next Y'$ where $X \imp X'$ and $Y' \imp
Y$. Thus by a subsumption step $T = S \setminus \{X \imp \next Y\}$. The proof is
similar to the above.
\end{proof}

\noindent We now introduce the concept of a {\em reduced behaviour
graph}, which will be used later in the completeness proof.

\begin{defn}{[Reduced Behaviour Graph]}
Given a behaviour graph we apply the following rules repeatedly until
no more deletions are possible.
\begin{itemize}
\item If a node has no successors, delete that node (and all edges to
      the node).  
\item If a node $n=(V,E)$ contains an eventuality $l$ (i.e.\ $l \in E$) and $l$ is not
      satisfied in 
      $n$, i.e.\ $l \not \in V$, and there is no path 
      from $n$ to a node whose valuation satisfies $l$, then
      delete $n$. 
\end{itemize}
The resulting graph is called the {\em reduced behaviour graph} for
$S$.
\end{defn}

\noindent This terminology implies that the reduced graph does not
depend on the order of deletions. The proof of this fact is
straightforward, but is not necessary for the completeness proof --- we
only need to know that {\em a} reduced graph (one from which no further
deletions are permitted) exists.
\begin{lemma}\label{cor1}
During the construction of a reduced behaviour graph
any node reachable from a deleted node is also deleted.
\end{lemma}
\begin{proof}
There are two conditions for the deletions of nodes to form a reduced
behaviour graph. Firstly nodes with no successors are deleted. No nodes are
reachable from a node with no successors hence the lemma
follows. Secondly nodes $n=(V,E)$ that are deleted where $l$ is an outstanding
eventuality, i.e.\ $l \in E$ but no reachable node satisfies $l$,
i.e.\ $\neg l \in V$. From the construction of the behaviour graph and from the
conditions allowing us to delete $n$, any node
$n'=(V',E')$ reachable from $n$ must contain $l$ as an outstanding
eventuality, i.e.\ $l \in E$ and  but doesn't satisfy $l$. Thus any node
reachable from $n$ must also be deleted.
\end{proof}
\begin{lemma}
\label{empty.unsat}
A set of SNF PLTL-clauses is unsatisfiable if, and only if, its reduced
behaviour graph is empty.
\end{lemma}

\begin{proof}
Let $S$ be a set of SNF PLTL-clauses. An infinite path through the
(unreduced) behaviour graph for $S$, starting at an initial node gives
a sequence of valuations for the propositional symbols --- i.e., a PLTL model. By
construction of the graph, this model satisfies the initial and step
PLTL-clauses of $S$. Furthermore, by Lemma~\ref{any.model} any such model
must arise from a path 
through the behaviour graph. However, not all paths give models for
the full set of PLTL-clauses $S$, since either the paths may not be infinite or
they may fail to satisfy some 
eventualities (which occur within sometime PLTL-clauses). If a node, $n$, has
no successors, then there are no infinite paths through that node, so
any model for $S$ must arise from a path through the graph with $n$
deleted. Thus the first deletion criterion can be applied without removing
any potential models. Also, if a node $n$ contains an eventuality $l$ then
any path 
through that node which is to yield a model for $S$ must satisfy $l$
either at $n$ or somewhere later in the path. 
% \fbox{\sf Unclear:} {\tt
Thus, if a node contains an eventuality that cannot be satisfied then this
node cannot be part of a model for the set of PLTL-clauses, hence, we can apply
the second deletion criterion without discarding 
potential models for $S$. The ``if'' part of the proposition follows.

To prove the ``only if'' part, suppose the reduced behaviour graph for
$S$, call it $G$, is non-empty. We will now use $G$ to construct a
model for $S$. First note that the set of initial nodes in $G$ is
non-empty, since, in the behaviour graph, every node is reachable from the
initial nodes and any node reachable from a deleted node is also deleted
(by Lemma~\ref{cor1}).
% since $G$ itself is non-empty and has been reduced. 
Now, choose an initial node $n_0 = (V_0,E_0)$. If $E_0$ is non-empty,
choose an ordering $e_1, \ldots,e_k$ for the literals in $E_0$. Since
$n_0$ has not been deleted, there is a path in $G$ to a node $m_{0,1}$
in which the eventuality $e_1$ is satisfied. If the eventuality $e_2$
is not present in $m_{0,1}$ it must have been satisfied somewhere
along the path. Otherwise, we can extend the path to a node $m_{0,2}$
which satisfies $e_2$. Continuing in this way we can find a path $P_1$
(which may consist simply of the node $n_0$ if all of $E_0$ are
satisfied there) such that each element of $E_0$ is satisfied at some
point along $P_1$. Let $n_1$ be a successor of the end point of $P_1$
(it must have a successor since we have deleted all terminal
nodes). Repeating our construction, we can find a path $P_2$ beginning
at $n_1$ along which all the eventualities in $n_1$ are satisfied. Let
$n_2$ be a successor of the end point of $P_2$. Repeat this
construction until $n_i = n_j$ for some $i > j$, which must happen
eventually since $G$ is finite. Let $Q$ be the path $P_{i+1} \ldots
P_j$. Then the path $P = P_1 P_2 \ldots P_{i} Q Q \ldots$ has the property
that, for each node in the path, each eventuality in that node is
satisfied at some node later in the path. To see this, recall that if
a node contains an eventuality $e$ but does not satisfy $e$, then $e$
is in the eventuality set of all immediate successors of $l$. So,
either $e$ is satisfied before we reach the next $n_r$ or $e$ is an
eventuality in $n_r$ and so is satisfied along $P_r$. Furthermore $P$
is obviously an infinite path. It follows by the construction of the
behaviour graph that the sequence of valuations given by $P$ is a
model for $S$.
\end{proof}

\noindent We are now ready to prove the completeness theorem for
propositional clausal temporal resolution.

\begin{theorem}[Completeness]
If a well-behaved augmented PLTL-clause set, $S$, is unsatisfiable then the
temporal resolution procedure will derive a refutation when applied to
$S$.
\end{theorem}

\begin{proof}
The proof proceeds by induction on the number of nodes in the
behaviour graph of $S$. 

First we consider the effect of simplification and subsumption rules on the
behaviour graph for a set of PLTL-clauses. 
Given a set of PLTL-clauses $S$ let the application of
simplification and subsumption rules to $S$ result in the set of
PLTL-clauses $S'$. By Lemma~\ref{simp.sump} the behaviour graph of $S$ is identical
to that of $S'$. 
% Thus the application of simplification and subsumption
% rules does not affect completeness.

If the behaviour graph is empty, then the set
of initial PLTL-clauses in $S$ is unsatisfiable. By the completeness of
classical resolution, we can use step resolution on the set of initial
PLTL-clauses to derive the empty clause.

Now suppose the behaviour graph $G$ is non-empty. By
Lemma~\ref{empty.unsat}, the reduced behaviour graph is empty and so
there must be a node which can be deleted from $G$. If $G$ has a
terminal node $n = (V,E)$, let $R$ be the set of step PLTL-clauses whose left
hand sides are satisfied by $V$. Then, having deleted the next operator, the
right-hand side of the PLTL-clauses 
in $R$ form an unsatisfiable set $L$ of propositional clauses. By
completeness of classical resolution again, there is a refutation of
$L$. Choosing an element of $R$ corresponding to each element of $L$,
we can ``mimic'' this classical refutation by step resolution
inferences to derive a step PLTL-clause
\begin{equation}
l_1 \land \ldots \land l_k \imp \next \lfalse \label{impf}
\end{equation}
where each $l_i$ is a literal which is satisfied by $V$. The temporal
resolution procedure allows us to rewrite PLTL-clause (\ref{impf}) as
\begin{eqnarray}
\lstart & \imp & \neg l_1 \lor \ldots \lor \neg l_k \label{impfr1}\\
\ltrue & \imp & \next(\neg l_1 \lor \ldots \lor \neg l_k). \label{impfr2}
\end{eqnarray}
By Lemma~\ref{add.rules}, adding PLTL-clauses (\ref{impfr1}) and
(\ref{impfr2}) (and any other resolvents derived along the way) to $S$
produces a PLTL-clause set $T$ whose behaviour graph $H$ is a subgraph of
$G$. ($H$ is in fact a proper subgraph, since $H$ has no node whose
valuation is $V$. If $n$ was an initial node it doesn't satisfy the
initial PLTL-clause (\ref{impfr1}) as $l_i \in V$ for $i=1 \ldots k$. If $n$ was a
non-initial node, as the left hand side $\ltrue$ is satisfied by every node
in $G$ the successor of any node must also satisfy $(\neg l_1 \lor \ldots \lor
\neg l_k)$. As we have $l_i \in V$ for $i= 0 \ldots k$ no edges can be
drawn to $n$ so $H$ does 
not contain $n$.) Furthermore, $T$ is
well-behaved since it has 
exactly the same models as $S$. By induction, $T$, and hence $S$, has
a refutation.

If $G$ does not have a terminal node, then it must contain a node $n =
(V,E)$ such that some eventuality $l \in E$ is not satisfied at any
node reachable from $n$. Let $N$ be the set of nodes reachable from
$n$. For each $n_i = (V_i,E_i) \in N$, let $R_i$ be the set of step
PLTL-clauses in $S$ whose left-hand sides are satisfied by $V_i$. 
Let
\begin{equation}
A_i \imp \next B_i \label{loopr}
\end{equation}
be an SNF$_m$ PLTL-clause that is the result of applying the SNF$_m$ merging
operation to the PLTL-clauses in $R_i$.
Note $A_i$ is the conjunction of the left hand side of the PLTL-clauses in
$R_i$ and $B_i$ is the conjunction of the right-hand sides of the PLTL-clauses in
$R_i$ (contained in the next operator) and $V_i$ satisfies $A_i$. Note
$A_i$ and $B_i$ are simply  classical propositional formulae.
Then each $B_i$ logically implies
$\neg l$ since none of the $V_i$ in $N$ satisfy $l$. Each $n_i \in N$ leads to a
node $n_j$ satisfying $B_i$ for some $i$. Thus $n_j$ must satisfy $B_i
\land l$ or $B_i \land \neg l$. By definition each successor of a node in
$N$ is also in $N$ (as $l$ is unsatisfied in all nodes reachable from
$n_i$). As $l$ is not satisfied by any node in $N$ we have $B_i \land l$ is
unsatisfiable and thus $B_i \imp \neg l$ is valid (in classical
propositional logic).

Also each $B_i$ logically
implies the disjunction of the $A_i$'s corresponding to the successors
of $n_i$. 
As each node $n_i \in N$
leads to a node $n_j =(V_j,E_j)$ that satisfies $B_i$. By definition $n_j
\in N$ and $V_j$ satisfies $A_j$. Thus $B_i \land \neg \bigvee_k A_k$ is
unsatisfiable. Hence $B_i \imp \bigvee_k A_k$.
Hence, we can use SNF$_m$ PLTL-clauses of the form (\ref{loopr}) in an
application of temporal resolution. Let $A$ be the disjunction of the
$A_i$. Then each $V_i$ satisfies $\neg l \land A$. For each node $n_i$
in $N$ either there is a PLTL-clause $C \imp \sometime l$ in $S$ and the
valuation at $n_i$ satisfies $C$, or for each predecessor $p_i$ of
$n_i$ the valuation at $p_i$ satisfies $w_l$.

Let $T$ be the result of adding the loop resolvents (\ref{wl4}), (\ref{wl5}) and
(\ref{wl6}) from \S\ref{augment}, and let $H$ be the behaviour graph for
$T$. Then $H$ has no nodes from the set $N$. So $H$ is a proper
subgraph of $G$ by Lemma~\ref{add.rules} and $T$ is well-behaved by
Lemma~\ref{sound}. Once again, it follows by induction that
there is a refutation for $S$.
\end{proof}
\subsection{Termination}
\begin{theorem}
The resolution algorithm will terminate.
\end{theorem}
\begin{proof}
Following the translation to normal form the set of PLTL-clauses is
augmented so no new proposition symbols are required during the proof. Hence we
have a finite number of proposition symbols. Further, there are a finite number of
right and left hand sides we may obtain as initial and step PLTL-clauses
modulo ordering of the conjunctions or disjunctions. Simplification rules
mean that the left or right hand sides cannot grow indefinitely. Note that
the number of sometime PLTL-clauses does not change. Thus step (3) of the
algorithm in \S\ref{alg.full.tr} either generates $\lstart \imp \lfalse$ and
terminates or we have tried to resolve each PLTL-clause with every other 
and obtained no new PLTL-clauses i.e.\ something that isn't in the set
already (modulo ordering of conjunctions/disjunctions).

The argument is similar for the termination of step 5. Having augmented the
set of PLTL-clauses with the new proposition symbols needed to translate
resolvents from temporal resolution into SNF, at some point no new
resolvents will be generated as we have a finite set of possible
PLTL-clauses. 
\end{proof}

%%%%%%%%%%%%%%%%%%%%%%%%%%%% SECTION STARTS.............
\section{Complexity}
\label{sec.complex}
We consider the increase in number of proposition symbols and PLTL-clauses
generated by the translation to SNF followed by consideration of the
complexity of the resolution proof method.
\subsection{Translation to the normal form}
We consider two aspects of the complexity of translating an arbitrary
formula to SNF in detail, namely the maximum number of SNF PLTL-clauses
generated from a formula of a given size, and the number of new
proposition symbols introduced. Note in this section we do not include the
new $w_l$ proposition symbols as we consider this to be part of the resolution
method itself. 

\subsubsection{Number of PLTL-clauses generated}
We define the length `$\sf{len}$' of a formula $A$ as follows.
$$
\begin{array}{rcl}
\len(\sometime l) & = & 1  \qquad\qquad\quad l \mbox{ is a literal}\\
\len(l_1 \lor l_2 \ldots \lor l_n) & = & 1 \qquad\qquad\quad l_i \mbox{ are
literals and } n \geq 1\\
\len(\const{}) & = & 1  \qquad\qquad\quad \const{}
\mbox{ is one of } \ltrue, \neg \ltrue, \lfalse \mbox{ or } \neg \lfalse \\
\len(\next(l_1 \lor l_2 \ldots \lor l_n)) & = & 1 \qquad\qquad\quad l_i \mbox{
are literals and } n \geq 1\\
\len(\neg \always A) & = & 1 + \len(\neg A) \\
 \len(\always A) & = & 1 + \len(A)\\
\len(\neg \sometime A) & = & 1 + \len(\neg A) \\ 
 \len(\sometime A) & = & 1 + \len(A)  \qquad\qquad\quad A \mbox{ not
a literal}\\
\len(\neg \next A) & = & 1 + \len(\neg A) \\
 \len(\next A) & = & 1 + \len(A)  \qquad\qquad A \mbox{ not a
                                  disjunction of literals}\\ 
\len(\neg (A \until B)) & = & 1 + \len(\neg A) + \len(\neg B) \\
 \len(A \until B) & = & 1 + \len(A) + \len(B)\\
\len(\neg (A \unless B)) & = & 1 + \len(\neg A) + \len(\neg B) \\
 \len(A \unless B) & = & 1 + \len(A) + \len(B)\\
\len(\neg (A \lor B)) & = & 1 + \len(\neg A) + \len(\neg B) \\
 \len(A \lor B) & = & 1 + \len(A) + \len(B)  \qquad A \mbox{ and }
B \mbox{ not disjunctions of literals}\\
\len(\neg (A \land B)) & = & 1 + \len(\neg A) + \len(\neg B) \\ 
 \len(A \land B) & = & 1 + \len(A) + \len(B)\\
\len(\neg (A \imp B)) & = & 1 + \len(A) + \len(\neg B) \\
 \len(A \imp B) & = & 1 + \len(\neg A) + \len(B) \\
% \len(\neg (A \iff B)) & = & 1 + \len(A) + \len(B) + \len(\neg A) +
% \len(\neg B) \\  
%  \len(A \iff B) & = & 1 + \len(A) + \len(B) + \len(\neg A) +
% \len(\neg B)\\
\end{array}
$$

\begin{lemma}\label{no.of.rules}
For any proposition symbol $x$ and PLTL formula $W$, 
%% not including the $\iff$ operator, 
the maximum number of PLTL-clauses, generated from the translation
of $\tau_1[\always(x \imp W)]$, denoted by ${\sf clauses}(\tau_1[\always(x
\imp W)])$, will be at most $11 \times 
\len(W)$, i.e.\
$$
{\sf clauses}(\tau_1[\always(x \imp W)]) \leqslant (11 \times \len(W))
$$
\end{lemma}
\begin{proof}
The proof is by induction on the length of $W$. The base case is where
$W$ has length 1, i.e.\ it has the form $\sometime l$, $l_1 \lor \ldots
\lor l_n$, $\ltrue$, $\lfalse$, $\next(l_1 \lor \ldots \lor l_n)$.  As
illustrated in \S\ref{tran.to.snf} $\tau_1[\always(x \imp \sometime
l)]$ produces one PLTL-clause, $\tau_1[\always (x \imp (l_1 \lor \ldots \lor
l_n))]$ produces two PLTL-clauses and $\tau_1[\always (x \imp \const{})]$
produces two PLTL-clauses (where \const{} is $\ltrue$, $\neg \ltrue$,
$\lfalse$ or $\neg \lfalse$) and $\tau_1[\always (x \imp \next (l_1
\lor \ldots \lor l_n))]$ produces one PLTL-clause. In each case if the number of
PLTL-clauses produced is $M$,
$$
M \leqslant (11 \times 1).
$$
For the inductive hypothesis we assume that the theorem holds for
formula of length $n$ and examine each case for length $n+1$.  Again,
by considering the proofs in \S\ref{tran.to.snf}, the
maximum number of PLTL-clauses from  removing any operator (or negated operator)
is 11 (from $\neg (A \unless B)$).
$$
\begin{array}{rcl}
{\sf clauses}(\tau_1[\always(x \imp \neg (A \unless B))]) & = & 11\! +\! 
     {\sf clauses}(\tau_1[\always(y\! \imp \neg A)])\! +\!
     {\sf clauses}(\tau_1[\always(z\! \imp \neg B)])\\
     & \leqslant &     (11  + (11 \times \len(\neg A)) + (11 \times \len(\neg B))) \\
      & = &     11 (1 + \len(\neg A) + \len(\neg B)) \\
      & = &      11 \times \len(\neg (A \unless B))
\end{array}$$
$$
\begin{array}{rcl}
{\sf clauses}(\tau_1[\always(x \imp (A \unless B))]) & = & 6 + 
     {\sf clauses}(\tau_1[\always(y \imp A)]) +
     {\sf clauses}(\tau_1[\always(z \imp B)])\\
     & \leqslant &     (6  + (11 \times \len(A)) + (11 \times \len(B))) \\
      & \leqslant &     11 (1 + \len(A) + \len(B)) \\
      & = &      11 \times \len(A \unless B)
\end{array}$$

$$
\begin{array}{rcl}
{\sf clauses}(\tau_1[\always(x \imp \always A)]) & = & 6 +\! 
     {\sf clauses}(\tau_1[\always(y \imp A)])\\
     & \leqslant &     (6  + (11 \times \len(A)))\\
      & \leqslant &     11 (1 + \len(\neg A)) \\
      & = &      11 \times \len(\always A)

\end{array}$$

$$
\begin{array}{rcl}
{\sf clauses}(\tau_1[\always(x \imp (\neg \always A))]) & = & 1 + 
    {\sf clauses}(\tau_1[\always(y \imp \neg A)]) \\
     & \leqslant &     (1  + (11 \times \len(\neg A))) \\
      & \leqslant &     11 (1 + \len(\neg A)) \\
      & = &      11 \times \len(\neg \always A)

\end{array}$$
The cases for the other operators are similar.
\end{proof}
\begin{theorem}
For any PLTL formula $W$, 
%% not including the $\iff$ operator, 
the maximum number of PLTL-clauses generated from the translation
into SNF will be  at most $1 + (11 \times \len(W))$, i.e\
$$
{\sf clauses}(\tau_0[W]) \leqslant (1 + (11 \times \len(W)))
$$
\end{theorem}
\begin{proof}
Let $W$ be a PLTL formula. To transform it into SNF we
apply the $\tau_0$ transformation i.e.\
$$
\tau_0[W] = \tau_1[\always(x \imp W)] \land \always(\lstart \imp x)
$$
 From Lemma~\ref{no.of.rules} we know the maximum number of PLTL-clauses from
$\tau_1[\always(x \imp W)]$ is $11 \times \len(W)$; hence, the maximum
number for the translation of $W$ is $1 + (11 \times \len(W))$.
\end{proof}

\subsubsection{Number of new proposition symbols generated}

\begin{lemma}\label{no.of.props}
For any proposition symbol $x$ and PLTL formula $W$, 
%% not including the $\iff$ operator, 
the maximum number of new proposition symbols generated from the
translation of $\tau_1[\always(x \imp W)]$, denoted by
$\props(\tau_1[\always(x \imp W)])$, will be at most $4 \times
\len(W)$, i.e.\
$$
\props(\tau_1[\always(x \imp W)]) \leqslant (4 \times \len(W))
$$
\end{lemma}
\begin{proof}
The proof is by induction on the length of $W$. The base case is where
$W$ has length 1, i.e.\ it has the form $\sometime l$, $l_1 \lor
\ldots \lor l_n$, $\ltrue$, $\lfalse$, $\next(l_1 \lor \ldots \lor
l_n)$. Each of these produces no new proposition symbols so as $0 \leqslant
(4 \times 1)$ we are done. For the inductive hypothesis we assume that
the theorem holds for formulae of length $n$ and examine each case for
length $n+1$. Again we examine some of the cases involved.
% The number of new propositions is denoted $\props()$.
$$
\begin{array}{rcl}
\props(\tau_1[\always(x \imp \neg (A \unless B))]) & = & 4\! +\!
\props(\tau_1[ \always(y\! \imp\! \neg A)])\! +\!
\props(\tau_1[\always(z\! \imp\! \neg B)])\\ 
     & \leqslant &     (4  + (4 \times \len(\neg A)) + (4 \times \len(\neg B))) \\
      & = &     4 (1 + \len(\neg A) + \len(\neg B)) \\
      & = &      4 \times \len(\neg (A \unless B))
\end{array}$$
$$
\begin{array}{rcl}
\props(\tau_1[\always(x \imp (A \unless B))]) & = & 3 + \props(\tau_1[\always(y \imp A)]) +
\props(\tau_1[\always(z \imp B)])\\
     & \leqslant &     (3  + (4 \times \len(A)) + (4 \times \len(B))) \\
      & \leqslant &     4 (1 + \len(A) + \len(B)) \\
      & = &      4 \times \len(A \unless B)
\end{array}$$
$$
\begin{array}{rcl}
\props(\tau_1[ \always(x \imp (\always A))]) & = & 2\! +\!
\props(\tau_1[\always(y\! \imp A)]) \\ 
     & \leqslant &     (2  + (4 \times \len(A)) ) \\
      & \leqslant &     4 (1 + \len(A) ) \\
      & = &      4 \times \len(\always A)
\end{array}$$

$$
\begin{array}{rcl}
\props(\tau_1[\always(x \imp (\neg \always A))]) & = & 1 +
     \props(\tau_1[\always(y \imp \neg A)])\\
     & \leqslant &     (1  + (4 \times \len(\neg A))) \\
      & \leqslant &     4 (1 + \len(A)) \\
      & = &      4 \times \len(\neg \always A)
\end{array}$$
The cases for the other operators are similar.
\end{proof}
\begin{theorem}
For any PLTL formula $W$,
%% not including the $\iff$ operator, 
the maximum number of new proposition symbols, $N$, generated from the translation
into SNF will be  at most $1 + (4 \times \len(W))$, i.e\
$$
N \leqslant 1 + (4 \times \len(W))
$$
\end{theorem}
\begin{proof}
Let $W$ be a PLTL formula. To transform it into SNF we apply the
$\tau_0$ transformation i.e.\
$$
\tau_0[W] = \tau_1[\always(x \imp W)] \land \always(\lstart \imp x)
$$
 From Lemma~\ref{no.of.props} we know the maximum number of new
proposition symbols from $\tau_1[\always(x \imp W)]$ is $4 \times
\len(W)$. Hence the maximum number for the translation of $W$ is $1 +
(4 \times \len(W))$.
\end{proof}

\subsection{Step Resolution}
Both forms of step resolution are essentially equivalent to classical
resolution, for example the derivation of $\next \lfalse$ on the right hand
side of a step PLTL-clause is essentially a classical resolution proof on the
clauses of the right hand side of (a subset of) the step PLTL-clauses.
The complexity of this phase of the method is
equivalent to the complexity of carrying out several classical resolution
proofs on
(simple translations of) the SNF PLTL-clauses. Indeed, one approach to the
practical mechanisation of step resolution has been to translate the
SNF PLTL-clauses in to a form suitable for a classical resolution theorem
prover~\cite{Dix97-ICTL}. 

\subsection{Temporal Resolution}
In order to consider the complexity of the temporal resolution phase, 
we describe a (naive) algorithm to find PLTL-clauses with which to apply the
temporal resolution operation.

\subsubsection{A Naive Algorithm for Loop Detection}
Given a set of $m$ step PLTL-clauses, $R$, and an eventuality $\sometime l$
 from the right-hand side of a sometime PLTL-clause, we carry out the
following.
\begin{enumerate}
\item Construct the set of merged-SNF PLTL-clauses for the SNF PLTL-clauses in
$R$, i.e.\ apply 
the merged-SNF operation in \S\ref{sub.snf} to
each set of PLTL-clauses in each member of the powerset of $R$ obtaining the
set of (SNF$_m$) PLTL-clauses, $R^*$.
\item Delete any PLTL-clause $X_i \imp \next Y_i$ in $R^*$ such that it is
not the case that $Y_i \imp \neg l$.
\item Delete any SNF$_m$ PLTL-clauses, $X_i \imp \next Y_i$ in $R^*$ such that it
is not the case that 
$$Y_i \imp \bigvee_j X_j$$
where $X_j$ is the left-hand side of PLTL-clause $j$ in $R^*$.
\item Repeat 3 until no more SNF$_m$ PLTL-clauses can be deleted.
\end{enumerate}

\subsubsection{Correctness of Naive Algorithm}
\begin{theorem}
Given a set of step PLTL-clauses $R$ and an eventuality $\sometime l$, there
is a loop in $\neg l$ within $R$ if, and only if, the above algorithm
outputs a non-empty set of PLTL-clauses $L'$.
\end{theorem}

\begin{proof}
Consider a loop $L$ in $\neg l$ formed from the set of PLTL-clauses
$R$. Let the disjunction of the 
left-hand side of the SNF$_m$ PLTL-clauses in $L$ be $X$. As $L$ is a loop the right-hand
side of each SNF$_m$ PLTL-clause in $L$ implies both $\neg l$ and $X$. Assume there are
$n$ SNF$_m$ PLTL-clauses in $L$. Each SNF$_m$  PLTL-clause (or an
equivalent SNF$_m$  PLTL-clause) in $L$ must be in the
set $R^*$ before deletions as $L$ has been made by combining PLTL-clauses in
$R$. 

We next consider the deletion of any SNF$_m$ PLTL-clause in $L$ from
$R^*$. Step 2 of 
the algorithm will not remove any of the SNF$_m$ PLTL-clauses in $L$ from
$R^*$ as it 
removes SNF$_m$ PLTL-clauses whose right-hand side do not imply $\neg l$
but, by assumption, each 
SNF$_m$ PLTL-clause in $L$ has a right-hand side that implies $\neg
l$. Assume we are 
about to remove a SNF$_m$ PLTL-clause $P \imp \next Q$, contained in $L$
from the set 
$R^*$ using step 3 of the algorithm. Let $Y$ be the disjunction of the
left-hand sides of the SNF$_m$ PLTL-clauses remaining undeleted in $R^*$
that are not 
in $L$. Thus $P \imp \next Q$ is being deleted as it is not the case
that $Q \imp X \lor Y$. However we know that $Q \imp X$, as $L$ is a
loop, so $Q \imp X \lor Y$ must also hold giving a
contradiction. Hence none of the SNF$_m$  PLTL-clauses in $L$ can be deleted from
$R^*$ so the algorithm must return a set of SNF$_m$  PLTL-clauses containing $L$.

Consider any set of SNF$_m$ PLTL-clauses $L'$ output by the algorithm. Each
SNF$_m$  PLTL-clause has been 
made by combining PLTL-clauses in $R$. Each right-hand side implies $\neg l$
otherwise it would have been deleted by step 2 of the algorithm. Each
right-hand side implies the disjunction of the left-hand side of the set
of SNF$_m$ PLTL-clauses otherwise it would have been deleted by step 3 of the
algorithm. The set of SNF$_m$ PLTL-clauses satisfies the side conditions for being a
loop, hence this loop can be constructed by combining the relevant PLTL-clauses in
$R$. 
\end{proof}

\subsubsection{Complexity of the Naive Algorithm}
Next we consider the complexity of detecting a set of PLTL-clauses in the way
outlined above. We assume a set of $m$ step PLTL-clauses containing $n$
proposition symbols.  The cost of combining the set of PLTL-clauses $R$ is $2^m$. To
check that the right-hand side of each PLTL-clause implies $\neg l$ 
% we just make sure that each disjunct on the right-hand side of each PLTL-clause
% implies $\neg l$. For each PLTL-clause, 
we must check a truth table with
$2^{n-1}$ lines. Thus for $2^m$ PLTL-clauses we must check in total $2^{n-1}
\times 2^m = 2^{m+n-1}$ lines.  For step 3 the worst case is if one
PLTL-clause is deleted from the set during each cycle of deletions until all
the PLTL-clauses are deleted. We must check each PLTL-clause implies the disjunction
of the remaining left-hand sides, i.e.\ for each right-hand side
checked we must consider a truth table with $2^n$ lines. Thus, to
check each PLTL-clause once has complexity of order $2^m \times 2^n =
2^{m+n}$, and to carry out $2^m$ rounds of checking we require $2^{2m
+n}$. Hence, the complexity of applying the resolution rule once is of
order $2^{2m +n}$.
\vspace{1ex}

\noindent 
%While the above gives a complexity bound for this naive
%algorithm, it is also the complexity bound for any loop-finding
%algorithm. 
This gives the worst case bound for any loop checking algorithm.
Refined approaches to finding loops only improve the
average performance~\cite{Dix96-CADE,Dix98:bfs}.

\subsection{Complexity of the Temporal Resolution Method}
We consider the complexity of the whole method by looking at the behaviour
graph used in the proof for completeness of temporal resolution. Assume we
have $n$ proposition symbols (including those added for augmentation see
\S\ref{mod.rulesets}) and $r$ eventualities. 
Deletions in the behaviour graph represent either a series of step
resolution inferences or a temporal resolution inference.

The deletion of a terminal node (and edges into it) corresponds to construction of a
PLTL-clause $A \imp \next \lfalse$, i.e.\ complexity of a classical resolution proof.
The deletion of a terminal subgraph (one or more nodes) with $p$ an unsatisfied
eventuality corresponds to temporal resolution (with complexity $2^{2m+n}$
for $m$ PLTL-clauses).
The worst case is if we have to delete each node separately i.e.\ the worst case
complexity is 
the number of nodes multiplied by, the maximum of the complexity of a temporal
resolution step  and the complexity of classical resolution,
plus the complexity of classical resolution (i.e.\ resolution between start
PLTL-clauses to finish the proof). 
Although the number of PLTL-clauses we have may change at each step, the worst case
number of PLTL-clauses is $2^{2n}$, i.e.\ $2^n$ possible left hand sides and $2^n$
possible right hand sides.
Recall that nodes in the behaviour graph are pairs $(V,E)$ where $V$ is a
valuation of the proposition symbols in the PLTL-clause set and $E$ is a subset of the
eventualities. Thus the number of nodes in the behaviour graph 
$2^n \times 2^r$ (where $r \leqslant 2n$), i.e.\ at worst $2^{3n}$. Thus complexity
is of the order $2^{3n} \times 2^{2^{2n+1}+n} = 2^{2^{2n+1}+4n}$.

We note that the complexity of satisfiablility for PLTL is PSPACE complete
\cite{SC82}. The complexity for the resolution methods in
\cite{AbadiManna85,CaFa84,Ven86} and the tableau method in \cite{Gou84} is
not discussed in the relevant papers, but the complexity 
for Wolper's tableau \cite{Wol83} is given as exponential in the length of
the initial formula.

%%%%%%%%%%%%%%%%%%%%%%%%%%%% SECTION STARTS.............
\section{Related Work}
\label{related}
We consider three resolution based approaches for PLTL (or similar
languages) and then several implemented methods for PLTL.

\subsection{Resolution Methods for PLTL}
\subsubsection{Venkatesh}
Venkatesh \cite{Ven86}  describes a clausal resolution method for PLTL for 
future-time operators including $\until$. 
First, formulae are translated into a normal form containing a restricted
nesting of temporal 
operators. The normal form is 
$$
\bigwedge_{i=1}^{n} c_i \land \bigwedge_{j=1}^{m} \always c_j^{'}
$$
where each $c_i$ and $c_j^{'}$ (known as {\em clauses}) is a disjunction of
formulae of the form 
$\next^k l$, $\next^k \always l$, $\next^k \sometime l$ or $\next^k(l'
\until l)$ (known as {\em principal terms}) for $l$ and $l'$ literals, $k \geqslant
0$ and $\next^k$ denoting a series of $k$ 
$\next$--operators. 

The clauses in the normal form therefore either apply to the first moment
in time or to every moment in time (those enclosed in a
$\always$--operator). Resolution proofs are displayed in columns separating
the clauses that hold in each state. To determine unsatisfiability, the
principal terms 
(except $\next^k l$) in each clause are 
{\em unwound} to split them into present and future parts. For example
the clause $F \lor \sometime l$ is replaced by $F \lor l \lor \next
\sometime l$ and similarly for $\always$ and $\until$. Next, classical style
resolution is carried out between complementary literals relating to the
present parts of the clauses in each column or state. Then, any clauses in a
state that contain only 
principal terms with one or more next operators are transferred to
the next state 
and the number of next operators attached to each term is reduced by
one. 
This process is shown to be complete for clauses that contain no eventualities.
Formulae that contain eventualities that are delayed indefinitely due to
unwinding are eliminated and this process is shown to be complete.

This system makes use of a normal form which at the top level is similar
to ours, i.e.\ there are clauses that relate to to first moment in time (as
do our initial PLTL clauses) and to every moment in time (as our step and
eventuality PLTL-clauses). Venkatesh uses renaming to remove any
nesting of operators, as we do here, to rewrite into the normal form. Thus,
as with our system, new propositions are introduced into the normal form. The
main difference is that Ventatesh does 
not remove the temporal operators $\always$ and $\until$.

Our initial step resolution can  be compared with the resolution of
complementary literals in the first state and step resolution is comparable
to resolution of complementary literals in other states.

The main difference is the treatment of eventualities. The system described
in this paper looks for sets of formulae with which to apply the temporal
resolution rule to generate additional constraints that must be
fulfilled. Venkatesh looks for persistent unfulfilled eventualities.
In many ways the Venkatesh system behaves like a temporal tableau
system~\cite{Wol83,Gou84} 
but classical resolution inferences are applied within states. Repeated
states containing persistent eventualities are identified and the
unresolved eventualities eliminated, similar to the check for
unsatisfied eventualities in temporal tableau. 

The overall approach to  the system described in this paper generates
constraints until we 
obtain a contradiction in the initial state $\lstart \imp
\lfalse$. Venkatesh's approach reasons forward carrying clauses that are
disjunctions of terms involving one or more next operator to the next
moment, having deleted a next operator. This forward reasoning approach
seems similar to the work on the executable temporal logics {\sc
MetateM} \cite{BFGOR96:book}.

\subsubsection{Cavalli and Fari{\~n}as del Cerro}
A clausal resolution method for PLTL is outlined in \cite{CaFa84}. 
The temporal operators defined in the logic include $\next$, $\always$, and
$\sometime$ but do not include $\until$.
The method described rewrites formulae to 
a complicated normal form and then applies a series of temporal resolution rules.

A formula, $F$, is said to be in 
Conjunctive Normal Form (CNF), if it is of the form
\begin{displaymath}
F = C_1 \land C_2 \land \ldots \land C_n
\end{displaymath}
where each $C_j$ is called a {\em clause} and is of the following form.
\begin{eqnarray}
C_j & = & L_1 \lor L_2 \lor \ldots \lor L_n \lor \always D_1 \lor \always D_2 \lor \ldots \lor \always D_p \nonumber\\
& \lor & \sometime A_1 \lor \sometime A_2 \lor \ldots \lor \sometime A_q \nonumber
\end{eqnarray}
Here each 
$L_i$ is a literal preceded by a string of zero or more $\next$--operators,
each $D_i$ is a disjunction of the same general form as the clauses 
and each $A_i$ is a conjunction where each conjunct possesses the same
general form as the clauses. 
The resolution operations are split into three types, classical operations,
temporal operations and transformation operations. The former applying the
classical resolution rule and classical logic rewrites, the latter two
required for manipulations of temporal operators.
For example a temporal operation is of the form
that $\always x$ and $\sometime y$ can be resolved if $x$ and $y$ are
resolvable and the resolvent will be the resolvent of $x$ and $y$ with a
$\sometime$--operator in front.

Formulae are refuted by translation to normal form and repeated application
of the inference rules. Resolution only takes place between clauses
in the context of certain operators outlined in the resolution rules.

The method is only similar to our method as it uses translation to a
clause form, although the normal form is much more complicated.
The rules required to rewrite formulae into the normal form depend on
temporal theorems  and classical methods. Renaming and the introduction of new
proposition symbols is not required. 

The temporal and transformation operations take account of the temporal
operators to make sure that contradictory formulae occur at the same moment in
time. In our system this is done by translating to the normal form followed
by initial and step resolution. Several operations are defined to deal
with eventualities, for example the temporal operation given above, whereas
we have just the one temporal resolution rule. The 
following complex transformation operation can be applied to an eventuality
and is required to deal with the induction between $\always$ and $\next$
$$
\begin{array}{c}
\Sigma_3(\sometime E, F) = E \lor \next E \lor \ldots \next^{n-1}E \lor
\Sigma_i (\sometime (\neg E \land \next \neg E \land \ldots \land
\next^{n-1}\neg E \land \next^{n}E), F)\\
\mbox{And if } E \lor \next E \lor \ldots \next^{n-1}E \mbox{ or }
(\sometime (\neg E \land \next \neg E \land \ldots \land
\next^{n-1}\neg E \land \next^{n}E), F)
\mbox{ is resolvable} \\
\mbox{ then } (\sometime E, F) \mbox{ is resolvable.}
\end{array}
$$
where $\Sigma_i$ denotes the further application of a classical, temporal or
transformation operation and $\next^{n-1}$ denotes a string of $n-1$ next
operators. 
The method is only described for a subset of the operators that we use, i.e.\
a less expressive logic. Further, the completeness proof is
only given for the $\always$, $\sometime$, and $\next$ operators. An
implementation of the method has been developed however it is not clear
when to apply each operation to lead towards a proof.

\subsubsection{Abadi}
Non-clausal temporal resolution systems are developed for propositional
\cite{AbadiManna85} and then
first-order temporal logics \cite{AM90} that are discrete and linear 
and have finite past and infinite future. The systems are
developed first for fragments 
of the logic including the temporal operators $\next, \always, \mbox{ and }
\sometime$  and then 
extended for $\next, \always, \sometime, \unless\footnote{Abadi denotes
  $\unless$, {\em unless}   (or {\em weak until}), as $\until$.} \mbox{ and
  } {\cal P}$. The binary operator ${\cal P}$ is known as {\em precedes}
where $u {\cal P} v = \neg ((\neg u) \unless v)$.

Because the system is non-clausal many simplification and inference rules
need to be defined. The resolution rule is of the form
\begin{displaymath}
A <u, \ldots, u>, B <u, \ldots, u> \longrightarrow A <\ltrue> \lor \, B <\lfalse>
\end{displaymath}
where $A <u, \ldots, u>$ denotes that $u$ occurs one or more times in
$A$. Here occurrences of $u$ in $A$ and $B$ are replaced with $\ltrue$ and
$\lfalse$ respectively. To ensure the rule is sound each $u$ that is replaced
must be in the scope of the same number of $\next$-operators, and must
not be in the scope of any other modal operator in $A$ or $B$, i.e.\ they
must apply to the same moment in time. 
Other rules such as distribution and modality rules allow the format of the
expression to be changed, for example the $\always$-modality rule allows
any formula $\always u$ to be rewritten as $u \land \next \always u$.

The induction
rule deals with the interaction between $\next$ and $\always$ and is of the
form 
$$ 
w, \sometime u \longrightarrow \sometime(\neg u \land \next(u \land \neg w))
\mbox{ if } \vdash \neg (w \land u).
$$
Informally this means that if $w$ and $u$ cannot both hold at the same time
and if $w$ and $\sometime u$ hold now then there must be a moment in time
(now or) in the future when $u$ does not hold and at the next moment in
time $u$ holds and $w$ does not.
Both systems are shown complete.
A proof editor has been developed for the propositional system with the
$\next, \always, \mbox{ and }
\sometime$ operators. 

As there is no translation to a normal form many rules need to be specified
to allow for every different combination of operators. The resolution rule
only allows resolution of formulae within the same number of next operators
and can perhaps be compared with our step resolution rule except, due to our
uniform normal form, our step resolution rule is much easier to
apply. Finally the rule that corresponds with our temporal resolution rule
is the induction rule. This rule can only be applied if a complex side
condition is checked.

Although a proof editor has been developed for the restricted propositional
system it seems unlikely that Abadi's system lends itself to a fully automatic
implementation. This is because of the large number of rules that may be
applied. Further, the induction rule requires a proof as a side condition
to its usage which will make automatic proofs difficult.  The
implementation of the induction rule is not discussed. The temporal
resolution rule we have described in this paper is also complex, 
however we have considered its implementation in
\cite{Dix96-CADE,Dix98:bfs} and developed a fully automatic prototype
theorem prover based on this.

\subsection{Implementations}
%
% Related proof methods for PLTL have already been mentioned in \S\ref{intro}; 
We now briefly mention several implementations
available for linear time temporal logics. The Logics
Workbench~\cite{lwb96}, a theorem proving system for various modal
logics
% (see for example \cite{Heuerding98}),
available over the web, has a module for dealing with logics such as
PLTL~\cite{Sch98}. The implementation of this module is based on tableau with an
analysis of strongly connected components to deal with eventualities.
A tableau-based theorem prover for PLTL, called DP, has
also been developed~\cite{Gou84}. Although not dealing with temporal logics,
tableau based methods are also used in FaCT~\cite{Horrocks98}, a
description logics classifier with a sound and complete subsumption algorithm.  
Finally, the STeP
system~\cite{STeP95}, based on ideas presented in
\cite{MannaPnueli92:book,MannaPnueli95:book}, and providing both model
checking and deductive methods for PLTL-like logics, has been used in
order to assist the verification of concurrent and reactive systems
based on temporal specifications.

%%%%%%%%%%%%%%%%%%%%%%%%%%%% SECTION STARTS.............
\section{Summary}
\label{sec.conc}
In this paper we have described, in detail, a clausal resolution
method for propositional linear temporal logic (PLTL), and have
considered its soundness, completeness, termination and
complexity. The method is based on the translation to a concise normal
form, and the application of both step resolution (essentially
classical resolution) and temporal resolution operations. Since
temporal logics such as PLTL are useful for describing reactive
systems, the resolution method has a variety of applications in
verifying properties of complex systems. We believe that this
resolution system can form the basis of an efficient temporal
theorem-proving system that can out-perform other systems developed
for such logics. However, there is still work to be done in order to
realise this.  

\subsection{Future Work}
\label{fut.work}
A prototype version of this system has been implemented in Prolog,
primarily to test the {\em loop search\/} algorithms required for the
temporal resolution rule~\cite{Dix96-CADE}. A more refined C++
version, known as {\sc Clatter\/}, is currently under
development. Both these systems utilise the fact that step resolution
is very similar to classical resolution and consequently use a
resolution theorem prover for classical logic, namely {\sc Otter}, to
implement this part of the system~\cite{Dix97-ICTL}.

The normal form used in this paper (SNF) has been extended to apply to
other logics such as branching-time temporal logics~\cite{BF97:ctl}
and multi-modal logics involving both a temporal and a modal
dimension~\cite{DFW98:JLC}. Much of our current work involves
extending the clausal resolution approach to a wider variety of
temporal and modal logics. In each of these logics, not only must a
version of SNF be defined, but specialised resolution operations must
be developed dependent on the properties of the logic in question.

Just as strategies for classical resolution have been successful in
improving efficiency, we aim to develop similar strategies for
temporal resolution. In particular, we are interested in the most
efficient way to apply the resolution operations in order to reduce
the number of resolution inferences that are made that do not
contribute towards finding a proof. The work described in \cite{DF98-TIME}
outlines preliminary steps in the definition of a temporal set of
support. The set of support strategy for classical resolution restricts the
number of resolution inferences that can be made. Inferences can only be
made where one of the clauses being resolved is from a subset of the full
clause set known as the set of support. Thus if we are asked to prove that
$B$ is a logical consequence of $A$ (or $A \vdash B$) in resolution we
would try show $A \land \neg B$ is unsatisfiable. To use the set of
support strategy the clauses derived
from $A$ are separated from those derived from $\neg B$, the latter being
put into the set of support. Thus resolution inferences between two clauses
derived from $A$ are avoided. We are also developing and
applying a modified resolution operation that can be used in a more
flexible way, and also can be used with strategies such as set of
support. Initial results can be found in~\cite{FisherDixon97-ICTL}.

Finally as efficient subsets of classical logic, such as Horn clauses,
have been investigated we hope to define restrictions on the normal
form that allow temporal resolution to be carried out more efficiently
and investigate the classes of problem these subsets correspond to.

\subsection*{Acknowledgements}
This  work was  partially supported by a SERC research grant GR/H/44646, an
EPSRC PhD Studentship and EPSRC Research Grants GR/K57282 and GR/L87491. We
would like to thank Howard Barringer, Graham Gough, Alexander Bolotov,
Ullrich Hustadt and Anatoli Degtiarev for their helpful comments and
suggestions about this work. The 
authors would also like to extend their gratitude to all the anonymous
referees for their hard work. Their dedication and time spent reviewing
this paper is much appreciated and has led to a greatly improved paper.

%
% \bibliographystyle{alpha}
% \bibliography{Abbreviations,ours,tl,languages,theorem-proving,automata,branching,knowledge,modal}
\newcommand{\etalchar}[1]{$^{#1}$}

\end{document}